\documentclass{aa}  
\usepackage{color}
\usepackage{natbib,twoopt}
\usepackage[hyphenbreaks]{breakurl}
\usepackage[breaklinks]{hyperref} 
\usepackage{hyperref} 
\usepackage{float}
\usepackage{graphicx}
\usepackage{siunitx}
\bibpunct{(}{)}{;}{a}{}{,}             %% natbib format for A&A and ApJ
\definecolor{cobalt}{rgb}{0.06, 0.2, 0.65}
\hypersetup{
  colorlinks,
  citecolor=blue,
  linkcolor=blue,
  urlcolor=cobalt,
  urlcolor=blue,
}
\makeatletter
  \newcommandtwoopt{\citeads}[3][][]{\href{http://adsabs.harvard.edu/abs/#3}%
    {\def\hyper@linkstart##1##2{}%
     \let\hyper@linkend\@empty\citealp[#1][#2]{#3}}}
  \newcommandtwoopt{\citepads}[3][][]{\href{http://adsabs.harvard.edu/abs/#3}%
    {\def\hyper@linkstart##1##2{}%
     \let\hyper@linkend\@empty\citep[#1][#2]{#3}}}
  \newcommandtwoopt{\citetads}[3][][]{\href{http://adsabs.harvard.edu/abs/#3}%
    {\def\hyper@linkstart##1##2{}%
     \let\hyper@linkend\@empty\citet[#1][#2]{#3}}}
  \newcommandtwoopt{\citeyearads}[3][][]%
    {\href{http://adsabs.harvard.edu/abs/#3}
    {\def\hyper@linkstart##1##2{}%
     \let\hyper@linkend\@empty\citeyear[#1][#2]{#3}}}
\makeatother
\usepackage{graphicx}
\usepackage{txfonts}
\newcommand{\skirt}{\mbox{\sc skirt}}

\newcommand{\eagle}{\mbox{\sc{Eagle}}}

\newcommand{\jwst}{\mbox{\it JWST}}
\newcommand{\flares}{\mbox{\sc Flares}}
\newcommand{\flare}{\mbox{\sc Flare}}
\newcommand{\um}{\SI{}{\micro\meter}}
\newcommand{\thesan}{\mbox{\sc Thesan}}
\newcommand{\tng}{\mbox{\sc Illustris-tng}}
\newcommand{\cmcubed}{\text{cm}^{-3}}

\defcitealias{10.1093/mnras/staa3360}{\flares\,I}
\defcitealias{10.1093/mnras/staa3715}{\flares\,II}
\defcitealias{Vijayan_2022}{\flares\,III}
\defcitealias{10.1093/mnras/stac1368}{\flares\,IV}
\defcitealias{roper2023flares}{\flares\,IX}

\begin{document}

   \title{First Light And Reionisation Epoch Simulations (FLARES) XVI}

   \subtitle{Size evolution of massive dusty galaxies at cosmic dawn from UV to IR}

   \author{Paurush Punyasheel
          \inst{1,2,3}\fnmsep\thanks{\email{paupun@dtu.dk}\\ \email{f20190184@goa.bits-pilani.ac.in} }
          \and
         Aswin P. Vijayan\inst{1,3,4}
          \and
         Thomas R. Greve\inst{1,3}
         \and
         William J. Roper\inst{4}
         \and
         Hiddo Algera\inst{5,6}
         \and \\
         Steven Gillman\inst{1,3}
         \and
         Bitten Gullberg\inst{1,3}
         \and
         Dimitrios Irodotou\inst{7}
         \and
         Christopher C. Lovell\inst{8}
         \and
         Louise T. C. Seeyave\inst{4}
         \and \\
         Peter A. Thomas\inst{4}
         \and
         Stephen M. Wilkins\inst{4}
          }

   \institute{DTU Space, Technical University of Denmark, Elektrovej 327, DK-2800 Kongens Lyngby, Denmark
         \and
             Birla Institute of Technology and Science, Sancoale 403726, Goa, India
        \and
             Cosmic Dawn Center (DAWN) 
         \and
         Astronomy Centre, University of Sussex, Falmer, Brighton BN1 9QH, UK
         \and
         Hiroshima Astrophysical Science Center, Hiroshima University, 1-3-1 Kagamiyama, Higashi-Hiroshima, Hiroshima 739-8526, Japan
         \and
         National Astronomical Observatory of Japan, 2-21-1, Osawa, Mitaka, Tokyo, Japan
         \and
         The Institute of Cancer Research, 123 Old Brompton Road, London SW7 3RP, UK
         \and
         Institute of Cosmology and Gravitation, University of Portsmouth, Burnaby Road, Portsmouth, PO1 3FX, UK
 }

   % \date{Received \today; accepted \today}

% \abstract{}{}{}{}{} 
% 5 {} token are mandatory
 
  \abstract
{We use the First Light And Reionisation Epoch Simulations (\textsc{Flares}) to study the evolution of the rest-frame ultraviolet (UV) and far-infrared (FIR) sizes for a statistical sample of massive ($\gtrsim10^{9}$M$_{\odot}$) high redshift galaxies ($z \in [5,10]$). Galaxies are post-processed using the \textsc{skirt} radiative transfer code, to self-consistently obtain the full spectral energy distribution and surface brightness distribution. We create mock observations of the galaxies for the Near Infrared Camera (NIRCam) to study the rest-frame UV (1500 \AA) morphology. We also generate mock rest-frame FIR (50 \um) photometry and mock ALMA 158 \um\ ($0.01''-0.03''$ and $\approx$$0.3''$ angular resolution) observations to study the dust-continuum sizes. We find the effect of dust on observed sizes reduces with increasing wavelength from the UV to optical ($\sim0.6$ times the UV at 0.4\um), with no evolution in FIR sizes. 
Observed sizes vary within $0.4-1.2$ times the intrinsic sizes at different signal to noise ratios (SNR = 5-20) across redshifts. The effect of point spread function (PSF) and noise makes bright structures prominent, whereas fainter regions blend with noise, leading to an underestimation (factor of $0.4-0.8$) of sizes at SNR=5. At SNR=15-20, the underestimation reduces (factor of $0.6-0.9$) at $z=5-8$ but due to PSF, at $z=9-10$, bright cores are dominant, resulting in an overestimation (factor of 1.0-1.2) of sizes. For ALMA, low ($\approx$$0.3''$) resolution sizes are affected by noise which acts as extended emission.
The size evolution in UV broadly agrees with current observational samples and other simulations. This work is one of the first to analyse the panchromatic sizes of a statistically significant sample of simulated high-redshift galaxies, complementing a growing body of research highlighting the importance of conducting an equivalent comparison between observed galaxies and their simulated counterparts in the early Universe.
}


   \maketitle
%
%-------------------------------------------------------------------

\section{Introduction}

Galaxy sizes can provide important insight into the physical processes responsible for galaxy formation and evolution. It is one of the few parameters which can be defined independently by photometry. Sizes in various wavelength ranges across various redshifts can not only tell us about evolution of galaxy properties with time \citep{Conselice2014}, but galaxies of various sizes and morphologies are important to constraint relations like the ultraviolet (UV) luminosity function \citep{Kawamata_2018} by providing completeness measure of data \citep{10.1093/mnras/stac380}. Observations have also shown that sizes of galaxies at fixed stellar mass are smaller at higher redshifts \citep{Franx_2008}, hence by studying the galaxy sizes over cosmic time we can study galactic processes affecting the structures of galaxies and thus galaxy evolution. This evolution in galactic processes can also change the dust and gas content, ultimately affecting the observed morphology. Galaxy mergers, instabilities, gas accretion, gas transport, star formation, and feedback can all affect the sizes of galaxies \citep{Xie_Li_Hao_2017}. With new high-resolution UV and far-IR imaging provided by \jwst\ and ALMA, the time is ripe for panchromatic studies of high-redshift galaxies.  
\\
\\
At a fixed redshift the relationship between luminosity and size can be expressed as 
\begin{equation}
\text{R} = \text{R}_0  \left( \frac{\text{L}_{\text{UV}}}{\text{L}^*_{z=3}}\right)^{\beta}
\label{eqn:lumpowerlaw}
\end{equation}
where $R_0$ is a normalization factor representing the effective radius at $\text{L}^*_{z=3}$, $\beta$ \footnote{$\beta$ used here is not equivalent to the UV spectral slope} is the slope of this relation, and $\text{L}^*_{z=3}$ is the characteristic rest-frame UV luminosity for $z \simeq 3$ Lyman break galaxies\footnote{This corresponds to $\text{M}_{1600} = -21.0$ mag or $L_{\nu} = 10^{29.03}\ {\rm erg\ s^{-1}\ Hz^{-1}}$ \cite[]{Steidel1999}}.

Studies \citep{Shibuya_2015,Holwerda_2015,Grazian,Bouwens_2022} also form the base of a growing consensus of a positive $\beta$ in equation \ref{eqn:lumpowerlaw}. However, the evolution of these values with redshift varies between studies. \citet{Shibuya_2015} finds a constant $\beta=0.27 \pm 0.01$ and a decreasing $R_0$ with redshift ($z<10$). \citet{Holwerda_2015} finds $\beta = 0.24 \pm 0.06$ at $z \backsim 7$ which decreases to $\beta = 0.12 \pm 0.09$ at $z \backsim 9-10$. \citet{Grazian} finds $\beta = 0.3-0.5$ at $z \backsim 7$. \citet{Huang_2013} states $\beta = 0.22$ at $z=4$ and $\beta = 0.25$ at $z=5$. Recent studies of Hubble Frontier Field lensed galaxies at $z=4-8$ agrees with a positive and high $\beta$ ($\backsim 0.4$) value \citep{Bouwens_2022}. They also show an increasing beta with increasing redshift, but the samples are limited to two bins of $z=4$ and $z=6-8$, which is insufficient to draw clear conclusions on the evolution of $\beta$.

There have also been a number of studies that have explored the size-luminosity relation and the associated physics influencing this relation at these high redshifts, using simulations of galaxy formation and evolution.
Studies such as \citep{10.1093/mnras/stw2912,10.1093/mnras/stz1810}, using semi-analytical models (SAMs) state $\beta=0.33$ for $5 \leq z \leq 10$ galaxies.
Many hydrodynamical simulations have also explored the evolution of observed galaxy sizes in the Epoch of Reionisation (EoR) \textbf{($z>5$)}. \citet{10.1093/mnras/stac380} studies the size-luminosity relations in the \textsc{BlueTides} simulation \citep{10.1093/mnras/stv2484}, which shows a decreasing positive $\beta$ with increasing redshift. 
Using the \textsc{Simba} simulations \cite[]{10.1093/mnras/stz937}, \citet{10.1093/mnras/staa1044} shows a similar positive $\beta$ for the FUV sizes. \citet{10.1093/mnras/stac1368} \citepalias[hereafter][]{10.1093/mnras/stac1368}, using the \textsc{Flare} simulations \cite[]{10.1093/mnras/staa3360,10.1093/mnras/staa3715}  also shows $\beta$ in the range of [0.279,0.319] for redshifts 5 to 8 and $\beta=0.519$ for redshift 9.

Galaxy size evolution as a function of redshift can be expressed as
\begin{equation}
\text{R}(z) = \text{R}_{0,z=0} (1+z)^{-m}
\label{eqn:Power Law}
\end{equation}
where $R_{0,z=0}$ is a normalization factor corresponding to the size of a galaxy at \textit{z} = 0 and $m$ is the slope of the redshift evolution \citep{10.1046/j.1365-8711.1998.01227.x}. 
% $R_{0,z=0}$ is a free factor.

A number of studies using deep \textit{Hubble Space Telescope (HST)} fields have measured the sizes of $z =
6-12$ Lyman-break galaxies \cite[e.g.][]{Oesch_2010,Mosleh_2012,Grazian,Ono_2013,Huang_2013,Holwerda_2015,Kawamata_2015,Kawamata_2018,Shibuya_2015}. These observations find that high-redshift galaxies are small and bright, compared to lower-redshift galaxies, with their half-light radius ($\text{R}_e$) in the range of $0.5-1$ kpc at rest-frame wavelength of 1500 \AA. 
These studies report slopes usually  in the range $0.6 \leq m \leq 2.0$ (equation~\ref{eqn:Power Law}), for galaxies with L$_{\rm UV}$ in the range $(0.3-1)$L$^{*}_{z=3}$. 
 
The size evolution with redshift have also been studied using galaxy simulations
studies, such as \cite{10.1093/mnras/stw2912,10.1093/mnras/stz1810}, using SAMs, predicting $m$ values of $1.9-2.2$ for L$_{\rm UV}$ in the range $(0.3-1)$L$^{*}_{z=3}$. These values are steeper than most observations,  which is accredited to strong supernova feedback in their model which produces a faster evolution of average galaxy size.
\citetalias{10.1093/mnras/stac1368}, find $m$ values of
$1.1-1.8$ for L$_{\rm UV}$ in the range $(0.3-1)$L$^{*}_{z=3}$.
These values are consistent with observational studies mentioned above. \textsc{BlueTides} simulations  \cite[]{10.1093/mnras/stac380} show a size evolution slope $m=0.662$, which is much lower than observations as well as other simulations, which may be the result of the simulations running only for $z\geq 7$, a regime which does not have extreme growth in galaxy size\textbf{.}

With various studies available, there is no clear consensus on the evolution of $\beta$ with redshift or $m$, in observational or theoretical studies. It is important to note that there are physical uncertainties in models, like the recipes for star formation and feedback which are used in simulations. \citet{Conroy_2009} also discusses  substantial uncertainties existing in stellar population synthesis (SPS) modelling. The choice of SPS models will affect the placement of a galaxy in the mass-/luminosity-size plane. However, the relative distribution of light in the galaxy will remain the same, leading to no size variation. Besides this both observational and simulation studies use different galaxy structure definitions as well as resolutions and observing instruments which results in variation of luminosity measurement. Simulations look for gravity-bound particles to define a galaxy whereas observations consider objects inside an aperture so the two datasets can differ in the structures that are identified as galaxies. Bleeding of light from nearby sources as well as background/dark sources can also play a role in the studies not being in agreement. Using similar analysis to compare models and observations becomes really important to constraint physical relations like size-luminosity relation.

Size evolution analysis is hard at EoR due to galaxies being hard to detect and resolve. 
% \jwst\ has started to provide a lot of FUV data for high-z galaxies. 
\jwst\ has changed this, due to it unprecedented sensitivity in the observed near-to-mid infrared.
However, in the far-infrared (FIR, rest-frame) ($z\ge5$) because of their lower dust content \citep{Magnelli_2020, refId0}, facilities such as ALMA require significant time investment to achieve meaningful observations. Thus, there are only a handful of studies that have examined the evolution of sizes in FIR wavelengths at $z\geq 5$. 
Using the ALMA-CRISTAL survey \citet{mitsuhashi2023almacristal} and \citet{ikeda2024almacristalsurveyspatialextent} evaluates the sizes of typical star-forming galaxies at redshifts, $z=4-6$, while studying the dust-obscured star formation. They find that dust continuum sizes at 158 \um\ are approximately 2 times as extended as UV sizes. Similarly, \citet{2022ApJ...934..144F} also finds dust continuum and [CII] emission to be extended with sizes larger than FUV sizes. \citet{2024A&A...686A.187P} also finds extended dust emission in FIR up to 3 kpc, with the dust continuum sizes bigger by a factor of 2 than the UV sizes.
\citet{Gullberg19} also compared K-band size with 870 \um\ sizes, finding 870 \um\ sizes being on average 2.2 times smaller than K-band sizes, albeit with majority of the sample at lower redshift ($z \lesssim 4$). The high-redshift ($z\ge5$) ALMA studies are opposite to the findings of \citet{Popping_2021} using the \tng\ 50 simulation \cite[]{Nelson2019TNG50,Pillepich2019TNG50} as well as \citetalias[][]{10.1093/mnras/stac1368}. They accredit this to clumpy gas distribution around the outskirts of galaxies or massive star-forming clumps within galaxies as major source of UV emission.

Dust in any astrophysical system will impact the radiation traveling through it by the scattering and extinction of photons, and re-emitting them at longer wavelengths. This makes the observed spectra of galaxies very different from the dust-free spectra \citep{li2008optical}. The dust content in the EoR is expected to be less than in the present universe \citep{10.1093/mnras/stz2684,Vijayan_2019,Magnelli_2020, refId0,yates2023impact}. 
However, even small amounts of dust can lead to significant attenuation in the UV \citep{2018MNRAS.477..552B,Liang,firedust}.
This absorbed energy is re-emitted in the IR, thus studying size evolution in the IR is equally important to understand the dust properties and its evolution in galaxies. 
The distribution and composition of dust in a galaxy can lead to variability in the amount of dust extinction, decreasing the observed luminosity for dusty regions \citep{Witstok2023,Markov,Shivaei}. \citet{FIRERachelCochrane} using \mbox{\sc Fire} simulations has also shown that heavy dust obscuration in certain line of sights can cause galaxies being unobservable.
The higher the wavelength, the lesser dust affects the radiation (\citet{10.1093/mnras/stac380}, \citetalias{10.1093/mnras/stac1368}). Since size measurement is based on the observed luminosity, this decrease in luminosity caused by dust will also bring variation in the observed size-luminosity or size-redshift evolution relations. Such size variation between mock observations and dust-free sizes has been reported in the \thesan\ simulations \citep{2024arXiv240208717S}, with median dust-free UV sizes being lower than dust attenuated sizes (factor of 1-0.33), with the differences increasing with stellar mass (Figure 16 in their paper).

Radiative transfer in astrophysical systems can be simulated to calculate the effect of dust and scattering of light away and into the line of sight to get realistic emissions at different wavelengths. This work uses the \textsc{skirt} radiative transfer code \citep{CAMPS201520,camps2020skirt} to post-process a sample of galaxies in the \textsc{Flares} suite of simulations (selection criteria is discussed in \S \ref{sec.selection}) and generate their Spectral Energy Distribution (SED) as well as the surface brightness profile in the UV and IR. \textsc{Flares} is chosen to get a statistical population of massive galaxies ($\gtrsim 10^{9}$M$_{\odot}$) in the EoR. We then use the results of this radiation transfer simulation to explore the size-luminosity relation and size evolution of galaxies in the epoch of re-ionisation in UV and FIR regime.
We also compare it with a similar analysis in UV, on  \textsc{Flares} galaxies with a Line of Sight (LoS) method \cite[as discussed in][]{10.1093/mnras/staa3715} in \citetalias[][]{10.1093/mnras/stac1368}. %\citet{2023ApJ...946...71C} using TNG50, made mock NIRCam observations with \textsc{skirt} for observed-frame 2\um\ $-$ 3.6\um\ at $3 \leq z \leq 6$ to analyse size evolution. Their study found size evolution slopes of m=1.26 and m=1.15 for F200W and F356W respectively, which are consistent with observational slopes. They also found observed sizes for massive galaxies to be larger than intrinsic stellar mass sizes at lower redshift (z=3,4), attributing it to mass being more compact than observable stellar light, predicting heavy dust obscuration. 
Such a panchromatic study of galaxy sizes is also very relevant and timely, given the wealth of multi-wavelength data expected in the next decade.
This analysis will enable us to study the disparity between the intrinsic galaxy sizes, to what can be observed by the best instruments available. 

The rest of this paper is structured as follows. In \S \ref{sec2}, we
detail the \textsc{Flares} simulations and choices made in \skirt\ while also detailing the methodology used to make synthetic photometry and mock observations. We also discuss methods of size calculation used. \S \ref{sec:noiseeffect} expands on the effect of noise and PSF which leads to variation in observation sizes against simulation sizes. In \S \ref{sec:UVsizes} we analyse the galaxy size evolution in UV against redshift, luminosity and mass. We also compare the size disparity between observations and simulations in this section. \S \ref{sec:IRsizes} discusses the IR galaxy size evolution with respect to redshift and mass. \S \ref{sec:pan_analysis} presents a panchromatic analysis between the UV and FIR spectrum sizes. 

We present our conclusions in \S \ref{sec:conc}. We use a Planck year 1 cosmology throughout this paper, corresponding to $\Omega_m = 0.307$, $\Omega_{\Lambda} = 0.693$, $h$  $=$  $0.677 \ $.

%--------------------------------------------------------------------
\section{Methodology} \label{sec2}
\subsection{The FLARE simulations}
The First Light And Reionisation Epoch Simulations \cite[\flares,][]{10.1093/mnras/staa3360,10.1093/mnras/staa3715} is a suite of zoom-in simulations of 40 regions chosen from a 3.2 cGpc a side dark matter only box. \flares\ has the same gas and dark matter mass resolution as \eagle\ ($m_{\rm g} = 1.8 \times 10^6$ \(\textup{M}_\odot\) and $m_{\rm dm} = 9.7 \times 10^6$ \(\textup{M}_\odot\) 
 respectively). These regions were re-simulated until $ z = 4.67 $ with full hydrodynamics using the AGNdT9 configuration of the \eagle\ \citep{10.1093/mnras/stu2058,10.1093/mnras/stv725} galaxy formation model \cite[see Table 3 in][]{10.1093/mnras/stu2058}. This configuration produces similar stellar mass functions to the reference \eagle\ model. 
This configuration gives more energetic, less frequent AGN feedback events, and better reproduces the gas mass fractions of low mass galaxy groups compared to the reference \eagle\ model. The resulting properties such as the UV luminosity function have been shown to converge to observational constraints in \citet{10.1093/mnras/staa3360,10.1093/mnras/staa3715}.

The regions are selected at $z = 4.7$ and have a radius of 14 cMpc/h, spanning
a wide range of overdensities, from $\delta = -0.479$ to $0.970$ \cite[See Table A1 in][]{10.1093/mnras/staa3360}. The regions are deliberately selected to contain a large number of extreme overdense regions (16) to obtain a large sample of massive galaxies. Two regions each were also selected at each overdensity, multiples of the standard deviation ($n\times\sigma$) derived by fitting a Gaussian function to the log(1+$\delta$) distribution, with 
% rms overdensity $\sigma$ , in the range $\sigma \in 
$n=[4, 3, 2, 1, 0.5, 0, -0.5, -1, -2, -3]$. Additionally, two mean density regions were chosen, to increase the sampled volume of these common regions, along with  the two most underdense regions ($\delta \approx -0.45$) in order to
cover the whole dynamic range. These total up to 40 regions. The simulations were run with a heavily modified version of PGADGET-3 an N-Body Tree-PM smoothed particle hydrodynamics (SPH) code \cite[same as \eagle, last described in][]{Springel2005}.  

The \flares\ regions are predominantly biased towards extreme over-density regions within the dark matter simulation box. This choice provides \flares\ with a statistically significant sample of massive galaxies in the early Universe, which are expected to be biased towards such regions \cite[see][]{Chiang2013,Lovell2018}. To remove bias to overdense regions and to ensure a representative sample, a weighting scheme \cite[described in \S 2.4 in][]{10.1093/mnras/staa3360} is used when calculating mean-median statistics.

%-------------------------------------- Two column figure (place early!)
   
\subsection{Galaxy selection}\label{sec.selection}
A Friends-Of-Friends algorithm \cite[FoF, ][]{1985ApJ...292..371D} is used to find bound groups in the simulations. Amongst these bound groups, galaxies in \flares\ are identified with the SUBFIND algorithm \citep{10.1046/j.1365-8711.2001.04912.x,2009MNRAS.399..497D}. This algorithm finds saddle points in the density field of a FoF halo to identify self-bound substructures. The most bound particle of these structures is denoted as the centre of our galaxies. Stellar mass of a galaxy in the simulation is defined based on the star particles present within a 30 physical kpc (pkpc) radius around this centre. The same selection criteria as in \cite{Vijayan_2022} \citepalias[hereafter][]{Vijayan_2022} is used to generate our dataset. We only include galaxies with $\ge1000$ star particles to ensure our sample is well resolved. This is so that the galaxies are well defined physically and also have enough data for Monte Carlo radiative transfer. This selection criteria leaves us with galaxies with stellar mass, $\text{M}_{\star} > 10^{9.12}$ \(\textup{M}_\odot\) and UV luminosity $\text{L}_{\rm UV,1500 \AA} > 10^{26.9}$ $\text{erg}/\text{s}/\text{Hz}$. See Table \ref{tab:sample} for the detailed distribution amongst mass bins.
\begin{table*}
    \centering
    \begin{tabular}{c|ccccc|c}
    \hline
        \textbf{z} & ~ &\multicolumn{3}{c}{\textbf{Mass Bins}($M\textsubscript{\(\odot\)}$)}& ~ &  \textbf{Total} \\ 
        ~ & $10^9-10^{9.5}$ & $10^{9.5}-10^{10}$&$10^{10}-10^{10.5}$& $10^{10.5}-10^{11}$ & $10^{11}-10^{11.5}$ & ~ \\ \hline \hline
        5 & 1384 & 1433 & 830 & 185 & 12 & 3844 \\ 
        6 & 688 & 729 & 300 & 54 & 1 & 1772 \\ 
        7 & 333 & 329 & 100 & 11 & 0 & 773 \\ 
        8 & 172 & 134 & 24 & 1 & 0 & 331 \\ 
        9 & 78 & 40 & 7 & 1 & 0 & 126 \\ 
        10 & 32 & 9 & 3 & 0 & 0 & 44 \\ \hline
        \textbf{Total}& 2184 & 2555 & 1667 & 456 & 28 & 6890 \\ \hline
    \end{tabular}
\caption{Distribution of galaxies across different redshifts and mass bins}\label{tab:sample}
\index{tables}
\end{table*}

\subsection{\skirt\ modelling}

In this work we use the \skirt\ \citep{CAMPS201520,camps2020skirt} radiative transfer code to construct rest-frame UV and far-infrared images of the \flares\ galaxies. \skirt\ is set up to simulate each galaxy in our selection. The setup is the same as described in \citetalias{Vijayan_2022}, with a brief description provided here. 
The SEDs of young stellar populations (age $\le10$ Myr) are modelled using the MAPPINGS III \cite[see][for more details]{MappingsIII} template. These were produced using the MAPPINGS III photo-ionisation modelling code, to model the nebular emission from young stars. The model assumes the STARBURST99 SPS code \cite[]{Starburst99} for the stellar tracks and a Kroupa IMF \cite[]{Kroupa2002}. 
The older populations (age $>10$ Myr) are modelled using the BPASS \cite[see][for more details]{BPASS2018} SPS code, assuming the Chabrier IMF.
We assume a \citet{Weingartner_2001} SMC type dust mixture to simulate dust emission and attenuation effects, with the effect of self-absorption by dust taken into account. We take into account the heating of dust by CMB, because at these redshifts ($z=5-10$), the CMB temperatures can be comparable to dust temperatures, affecting the observed luminosities. Furthermore, as all these radiative transfer results are estimated by \skirt\ using Monte Carlo Method, a higher number of photons simulated would lead to a more accurate result, hence we use $10^6$ photon packets per radiation field wavelength grid (Appendix A in \citetalias{Vijayan_2022} shows that increasing the photon count has negligible effect on radiative transfer results with obtained SEDs by using $10^6$ and $10^7$ photon packets per wavelength grid converging). Other factors such as the distribution of sources, dust and gas along with dust mass fraction are defined for each galaxy based on the \flares\ star and gas particle data, similar to \citetalias{Vijayan_2022}. 

The observational instruments in \skirt\ are setup at a distance of 1 Mpc from the galaxy and they have a field of view (FoV) of 60x60 pkpc$^2$. This whole field of view is captured in a 400 $\times$ 400 pixels frame (this corresponds to a per pixel resolution of 0.023", 0.026", 0.028", 0.030", 0.033", 0.035" for redshifts 5 through 10 respectively)\footnote{This lies below the gravitational softening length ($s=2.66/(1+z)$ pkpc) of the simulation (As calculated later in \S \ref{sizecalculation}, our UV intrinsic sizes lie above the softening length but as IR sizes go below the softening length they can be unreliable).}. The plane on which these galaxies are observed is the same and hence the orientation of the galaxies in \flares, automatically leads to different viewing angles. Only at the highest redshifts ($z\ge9$) corresponding to the most massive galaxies (M$_{\star}\ge 10^{10}$ M$_{\odot}$), where there is a dearth in the number of galaxies sampled in \flares\ (see Table \ref{tab:sample}) and corresponds to a high disc galaxy fraction \cite[see][]{Ferreira2023}, there will be an increased effect of viewing angle (face-on/edge-on). \footnote{We ignore data bins that are not sufficiently populated (N<20) in all our size-evolution analysis. This reduces the effect of viewing angles. Our sample selection also encompasses a wide range of environments. With galaxies being selected from different regions, they are statistically randomly sampled, avoiding the intrinsic alignment of galaxies in nearby environments, as seen in simulations like \eagle\ \citep{HillEAGLE}.} 
Based on these parameters, we get a per pixel SED for both UV and IR spectrum as well as the SED for the range [0.08,1500] $\mu$m for the whole FoV. 
The images are centred on each galaxy's centre of potential as calculated by \textsc{subfind}.

\subsection{Comparison with Line of Sight method of previous studies}

While in this work we employ \skirt\ to produce synthetic images in most previous \flares\ work we have adopted a simpler approach to produce synthetic observations, including spectra and imaging. Line of Sight \cite[LoS, as discussed in][]{10.1093/mnras/staa3715} method estimates the effect of dust attenuation by calculating the 
intervening column density of dust. This is converted to an optical depth measure by assuming a dust extinction curve. Anisotropic scattering by the dust couples all lines of sight, and dust absorption/emission couples all wavelengths, making the radiative transfer equation highly non-local and nonlinear \citep{Steinacker_2013}, which is not taken into account in the LoS method, because of its complexity. However, the star-dust geometry is preserved. Radiative transfer using Monte Carlo simulation estimate these effects in both absorption and scattering by dust grains, and can self-consistently generate the dust emission in the IR, which is the major motivation for using \skirt\ outputs in this study. 
% Figure \ref{fig:lumdiff} which plots the radiative transfer luminosities which are $\approx 0.2-0.3$ dex lower than plotted LoS luminosities and shows the added effects which are not accounted for in LoS but also bring a variation in luminosities observed using both methods. 
Figure 1 shows the radiative transfer luminosities, which are $\approx 0.2-0.3$  dex lower than the line-of-sight (LoS) luminosities. It also illustrates the additional effects that are not accounted for in the LoS method but contribute to the observed variation in luminosities between the two approaches.
Also, in previous studies, in the LoS method, the free parameters that are used to calculate the dust attenuation were chosen to match the UV luminosity function at $z=5$ \cite[see \S 2.4 in][]{10.1093/mnras/staa3715}.
This difference in method of calculating the dust attenuated luminosity will cause the intercept and slopes of the size-luminosity relation to differ in the two methods. \S \ref{sec:sizelum} shows this difference aptly when compared with slopes and intercepts found in the Appendix B of \citetalias[][]{10.1093/mnras/stac1368}. Appendix E in \citet{Vijayan_2022} also compares the UV luminosity function obtained by using the LoS method and radiative transfer for the same sample of galaxies as this study. \citet{Vijayan_2022} also finds the normalization of radiative transfer results to be 0.4 dex lower than LoS model attributing it mainly to the lower dust optical depth along the line of sight adopted in LoS studies of \flares\ galaxies compared to the radiative transfer method.

\begin{figure}[]
\centering
   \includegraphics[width=8cm]{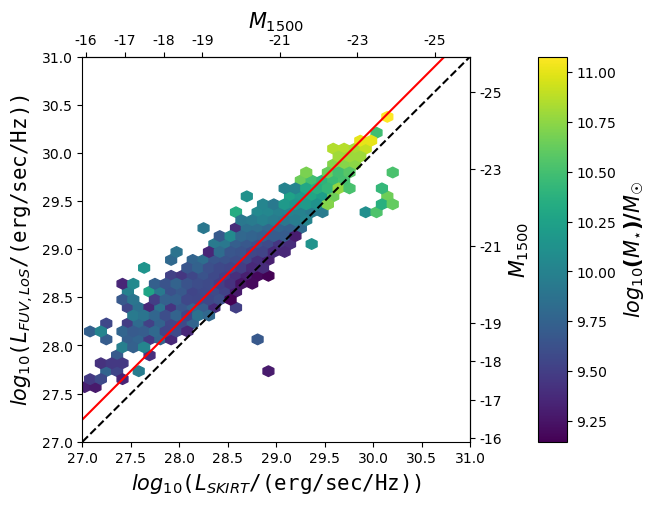}
     \caption{Luminosity from \skirt\ radiative transfer (estimated at $1500 \AA$) on x axis is compared with LoS luminosities \cite[as discussed in][]{10.1093/mnras/staa3715} on y axis. Luminosity from the LoS method is roughly 0.25-0.3 dex higher than luminosities produced by radiative transfer. The red line is the best fit offset between the two luminosities. The colours corresponding to the colourbar shows the average stellar mass in each hexbin\textbf{.} }
     \label{fig:lumdiff}
\end{figure}

\subsection{Size calculation} \label{sizecalculation}
Generally to measure sizes, observational studies use either S{\'e}rsic profile fitting \citep{1963BAAA....6...41S} or a curve of growth method \citep{Stetson_1990,Ferguson_2004, Bouwens_2004,Oesch_2010} based on circular or elliptical apertures. S{\'e}rsic profile fitting causes variation in sizes compared to the curves of growth sizes in case of a clumpy nature of high redshift galaxies \citep{Jiang_2013,10.1093/mnras/stab3744}. Clumpy galaxies are elongated in nature with multiple separated bright parts, amongst which centre selection plays an important role in half-light radius calculation, affecting the size calculated.  In simulations, particle distributions can also be used to find the radius enclosing half of the light or mass to determine sizes. Since our aim is to simulate observations and calculate their disparity with observed sizes we use circular apertures, which has been widely used in previous studies \citep{Bouwens_2004,Oesch_2010} and use curves of growth (luminosity contained inside a radius) to define our half-light radius. We use two different methods to do so which are described below. Previous \flares\ work \citepalias{10.1093/mnras/stac1368} have also used a non-parametric pixel based method \cite[e.g.][]{10.1093/mnras/stw3296} to calculate sizes. A comparison of the two methods is also presented below. 
\\ \\
\noindent \textbf{Iterative aperture method} \label{IterAp} \\
We calculate the total luminosity in the galaxy image by summing all pixels within the FoV. For calculating sizes (half-light radius) using the Iterative Aperture method, we start with the centre of potential of the galaxy. We calculate total luminosities in circular apertures of given radius R (starting with R=0) from the potential centre in increments of 1 pixel until the flux inside the aperture exceeds the 3/4th the total luminosity of the FoV. Then using the radii and their corresponding luminosity we interpolate the half-light radius (radius with half the total luminosity of the image) converted to pkpc. 
\\\\
\noindent \textbf{\texttt{statmorph}} \\
\texttt{statmorph}\footnote{\href{https://statmorph.readthedocs.io}{https://statmorph.readthedocs.io}} \citep{10.1093/mnras/sty3345} is an open source \texttt{python} package which segments an image to look for objects and calculates their morphological properties. It provides the centre of the object, its radius at various light fractions, concentration, asymmetry, clumpiness and Gini-M20 statistics. In this study we use \texttt{statmorph} on our mock images to get morphological properties for the galaxies if they were observed at different SNRs as discussed in sub\S \ref{nircam}.
\\

Iterative Aperture Method is used to calculate sizes for high resolution, noise and PSF free SKIRT outputs. These sizes will be denoted as "intrinsic sizes" or "$R_{e,intrinsic}$" throughout the following paper as these denote the size of galaxies with no observational effects at play. Iterative Aperture Method is used for these outputs as clumpy or dispersed galaxies with few connecting pixels are segmented as separate objects while using \texttt{statmorph}.

\begin{figure}
\centering
   \includegraphics[width=8cm]{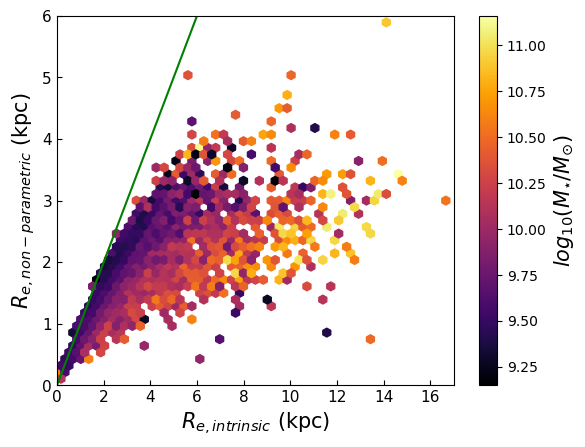}
     \caption{Comparison of the half-light radii at 1500 \AA\ using the method from \citetalias{10.1093/mnras/stac1368} and this studies' circular aperture method is shown above. The calculated HLR from a non-parametric pixel-based method is plotted on the y-axis with the intrinsic sizes plotted on the x-axis. Sizes from non-parametric pixel method are lower than intrinsic sizes used in this study due to non-parametric pixel method accounting for the most compact configuration of the galaxy. The green line shows the 1:1 relation }
     \label{fig:radiusroper22}
\end{figure}

\texttt{statmorph} is used to calculate sizes for all simulated imaging which contains observational effects. These sizes will be denoted as "observed sizes" or "$R_{e,obs}$" throughout the following paper. Appendix \ref{converge} shows that for noise and PSF free SKIRT output that can be segmented properly, the sizes calculated by both these methods converge to a 1:1 relation.
\\\\
\noindent \textbf{Comparison with non-parametric pixel method} \label{nonparam}\\
\citetalias{10.1093/mnras/stac1368} used a non-parametric pixel approach to calculate the half-light radius (denoted by $R_{e,non-parametric}$). In this method the pixels of the image are ordered from most luminous to least luminous, and then the pixel area containing half the total luminosity is taken into account to calculate the half-light radius assuming a circular aperture. This method is used to account for the clumpy nature of galaxies at high redshift. As this method gives the radius assuming the most compact morphology possible, the sizes calculated by our curves of growth method using circular aperture are higher as they are dependent on structure, clearly seen in Figure \ref{fig:radiusroper22} which plots the sizes from pixel method against the sizes from curves of growth method for 1500 \AA.

\begin{figure*}
\centering
   \includegraphics[width=17.5cm]{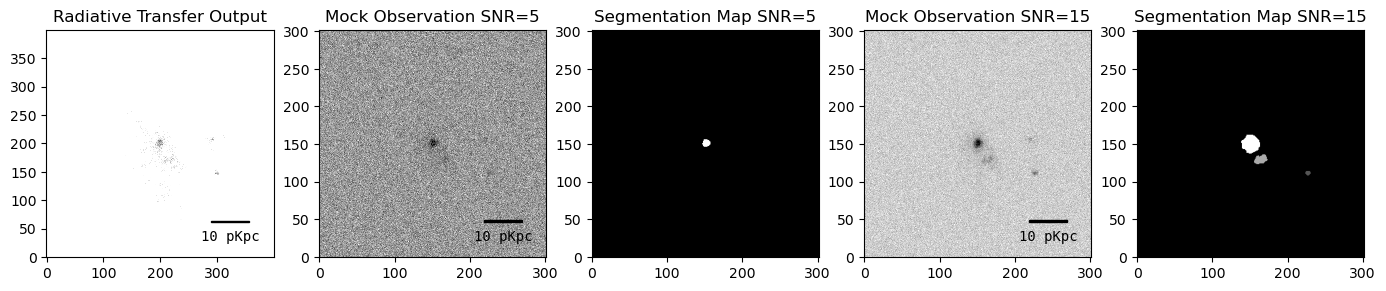}
     \caption{The panes show image of a sample galaxy at $z=5$ at various stages of methodology, in a 60x60 pkpc$^2$ field of view. First pane: The result of \skirt\ radiative transfer simulation. Second and Third Pane: Mock observation produced after regridding to 302x302 pixels based on  Near Infrared Camera (NIRCam) resolution at this redshift and adding the effects of Gaussian + shot noise corresponding to SNR=5 and PSF is shown in the second pane with its respective segmentation map in the third pane showing only a single detected object. Fourth and Fifth Pane: Simulated Imaging as described earlier but at SNR=15 is shown in fourth pane with its respective segmentation map showing three distinct objects detected in fifth pane}
     \label{fig:mockimage}
     
\end{figure*}
\begin{figure*}
\centering
   \includegraphics[width=17cm]{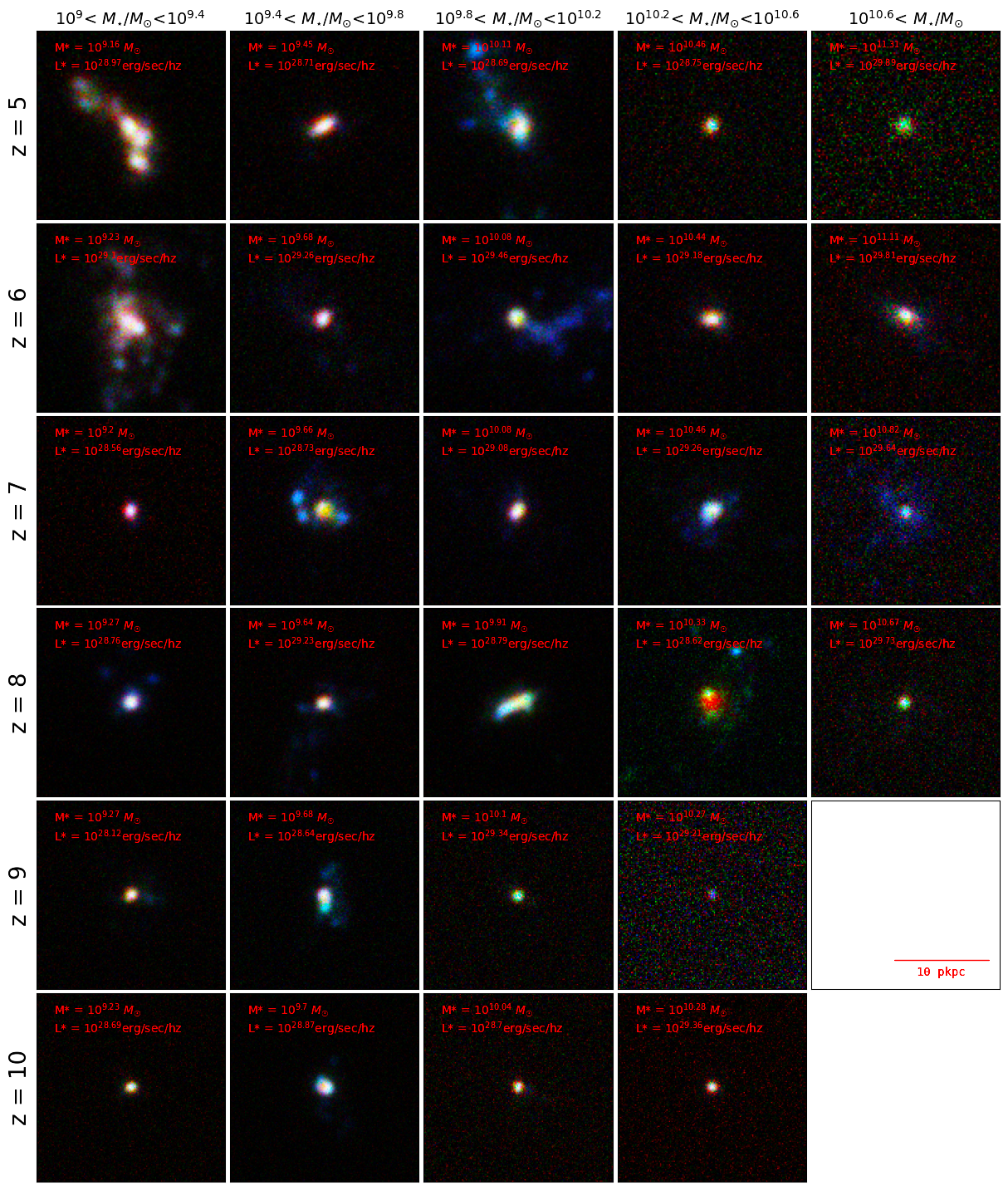}
     \caption{The above false colour images show sample galaxies at different redshifts and various mass bins at SNR=20. The field-of-view of the images is 20pkpc and the red, green and blue channel taken from JWST/NIRCam F200W, F150W, F115W filter data respectively.}
     \label{fig:NIRCamfalse}
\end{figure*}

It is important to note that for a direct comparison between \citetalias{10.1093/mnras/stac1368}, resolution also plays a role, as contrary to using the softening length ($s=2.66/(1+z)$ pkpc) resolution, this study uses fixed 400x400 pixel resolution for intrinsic sizes irrespective of redshift. All UV observational analysis in the study uses \jwst\ resolutions as described in \S \ref{nircam}.

\subsection{Photometry/image creation}
\subsubsection{NIRCam} \label{nircam}
For generating our mock observations, we use a rest-frame wavelength of 1500 $\AA$, corresponding to the far-UV. We make mock observations for \jwst\ NIRCam's filters: F090W, F115W, F140M, F150W, F162M, depending on the redshift. 
In the wavelength range of 0.6–2.3 \um\ observations, the NIRCam resolution is 0.031"/pixel. 
Using the field of view from \skirt\ and distance (i.e. redshift) we regridded the images produced from \skirt\ to the NIRCam resolution (see Table \ref{tab:resolution}).

To mock the observational effects of NIRCam observations, the images were convolved with a Point Spread Function (PSF) for different filters corresponding to the observed (redshifted) wavelength. We used the PSFs from \jwst\ PSF simulation library, simulated by WebbPSF \citep{2012SPIE.8442E..3DP,2014SPIE.9143E..3XP}, provided by STScI. To the convolved images we added shot/Poisson noise. To account for the sky background noise, the images were also further modified with Gaussian noise after PSF convolution. To analyse the effect of noise mathematically,  noisy mock image was generated uniquely for each source at different SNRs (5-20). Figure \ref{fig:mockimage} shows the evolution of image through various stages described above and Figure \ref{fig:NIRCamfalse} shows a sample of NIRCam RGB false colour images (F200W, F150W, F115W) for galaxies in different mass bins. We then evaluated results at Signal to Noise Ratios (SNR) of 5, 10 , 15, and 20 
(corresponding to exposure times of approximately 5, 14, 40, 70 minutes at $z=5$ for an observations similar to CEERS \citep{2024MNRAS.529.1067H} for a representative galaxy with NIRCam (F090W) flux), to see the effect of observation time on the observed sizes. 

 We use \textit{detect\_sources} from the \texttt{photutils} \citep{larry_bradley_2023_7946442} library to create a segmentation map (sources are defined with 1.5 $\sigma$ detection and a criterion of 5 minimum connected pixels). Depending on the background SNR added to the image, our segmentation maps can differ as seen in Figure \ref{fig:mockimage} which shows that three separate objects are detected at SNR=15 whereas at SNR=5 two of these objects blend with noise and aren't segmented.
 The segmentation algorithm which takes into account the minimum number of connected pixels for an object, hence can identify two or more objects from a single galaxy image. In these cases, we considered the brightest source as our galaxy. This also resolves cases of mergers with the two distinct cores being segmented as separate objects and only one of them being selected for size analysis. We also evaluated our galaxies, by simulating observations for being a part of a large survey with each galaxy embedded in a Gaussian noise field with a fixed standard deviation for all the sources. The effect of such an analysis is present in Appendix \ref{sec:noisefield}. For all further analysis results of SNR=5 imaging are used unless explicitly stated otherwise.

\begin{figure*}
\centering
   \includegraphics[width=17cm]{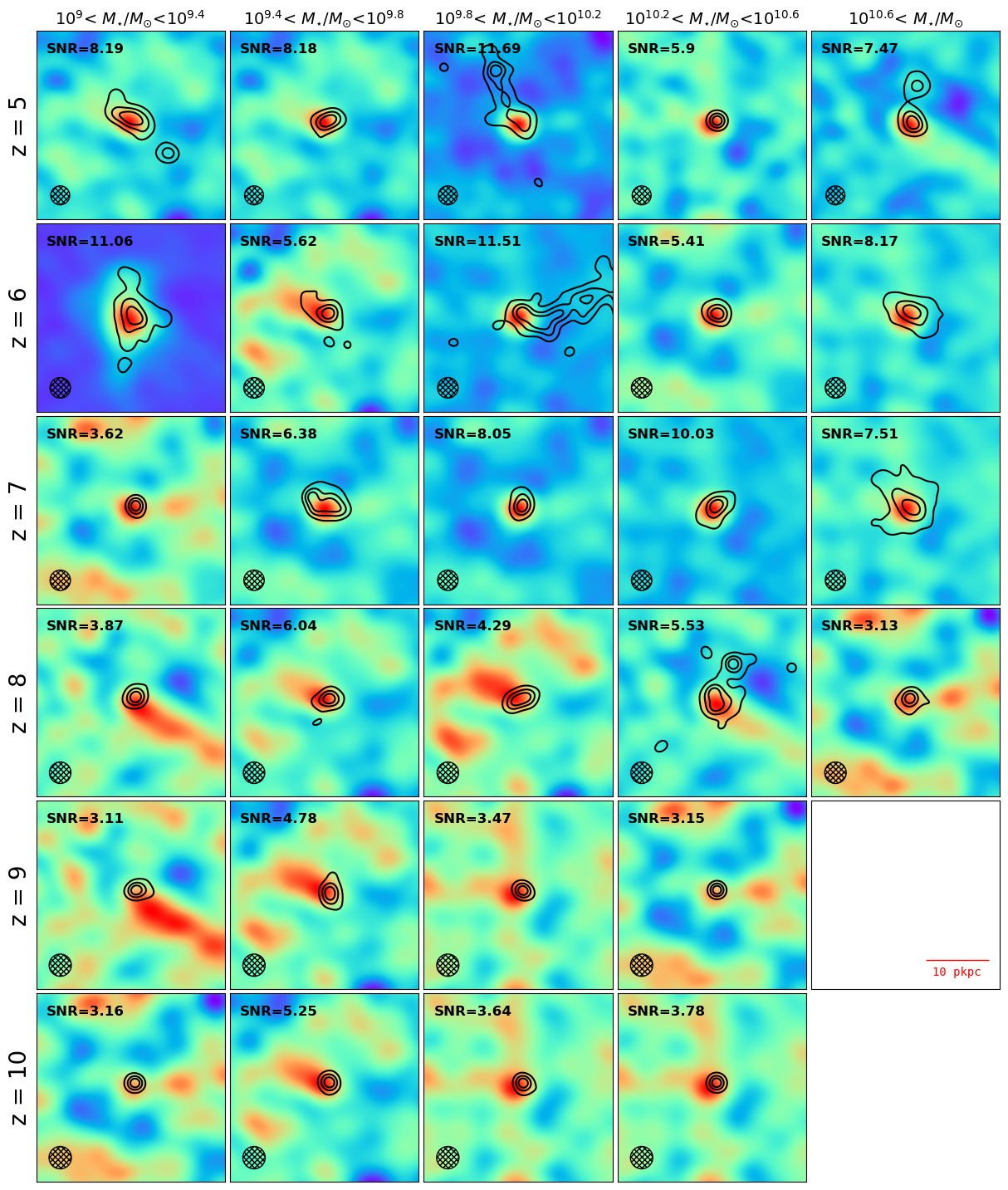}
     \caption{The figure shows the sample of galaxies (same galaxies as in Figure \ref{fig:NIRCamfalse}) as observed by ALMA at 158 \um\ at an aperture of 30 pkpc. Low angular resolution ($\approx 0.1"$) (similar to CRISTAL program) is used. The beam size is shown by the circle at bottom left corner. The black contours show 4 linearly spaced contours between brightest pixel to 1/4th of brightest pixel for 1500 \AA.}
     \label{fig:ALMAfalse}
     
\end{figure*}

\begin{figure*}
\centering
   \includegraphics[width=17cm]{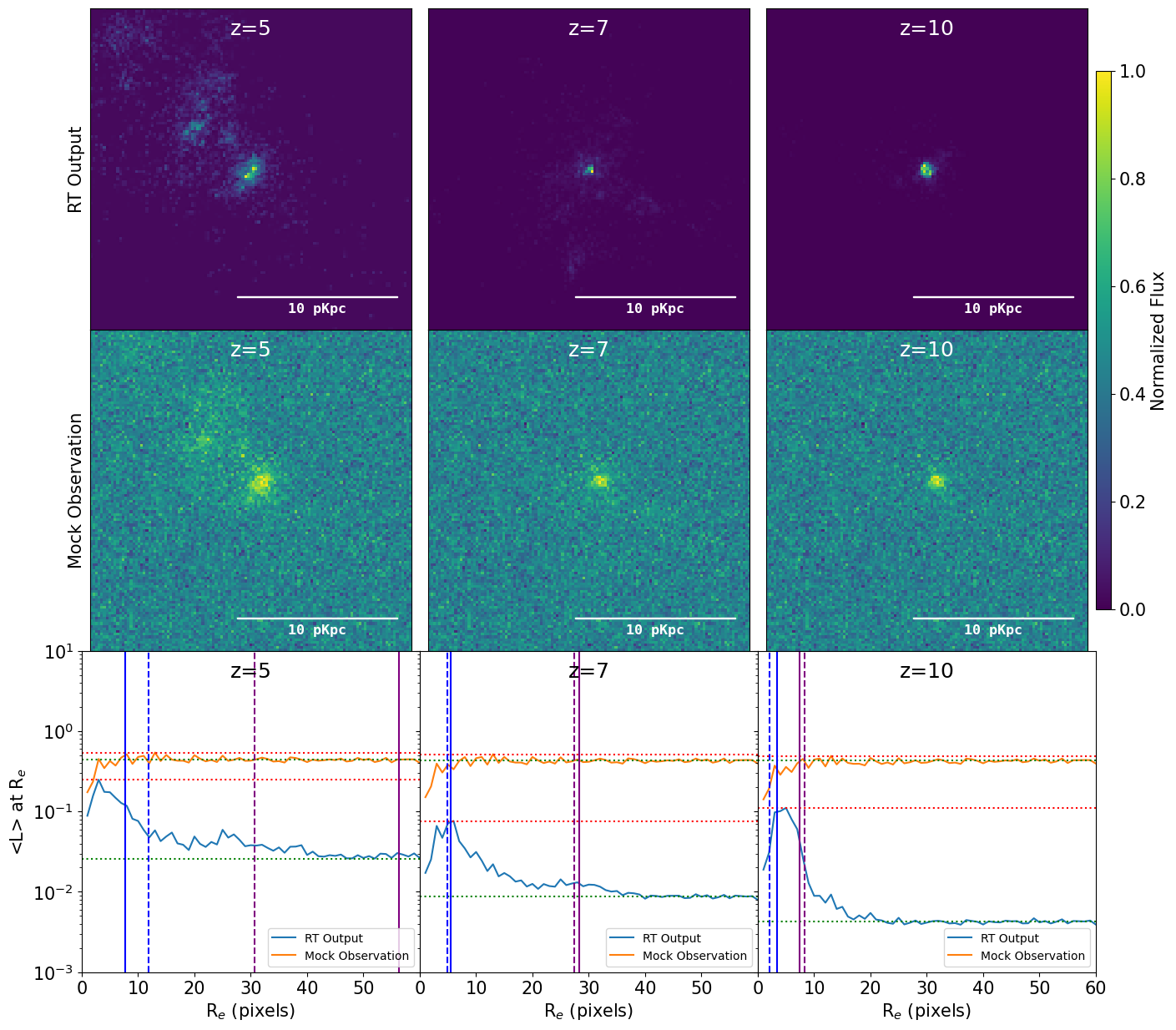}
     \caption{Brightness comparison between original and noisy image is shown for a sample redshift 5,7 \& 10 galaxies with size offset at SNR=5. The bottom row shows the mean luminosity of pixels at each radius (y-axis) calculated by finding pixel luminosities at the edge of the aperture and dividing it by circumference) plotted against the radius (x-axis) for the normalized images. The centre for these luminosities is assumed as the centre of potential as defined in the simulations.  The red lines indicate the maximum brightness and the green line shows the median brightness. The blue solid and dashed lines show the r20 of original and noisy image respectively. The purple solid and dashed lines show the r80 of original and noisy images respectively. Panel for z=5 shows the case of underestimation, z=7 of near parity and z=10 shows overestimation. }
     \label{fig:brightness comparision10}
     
\end{figure*}
\begin{figure*}
\centering
   \includegraphics[width=17.5cm]{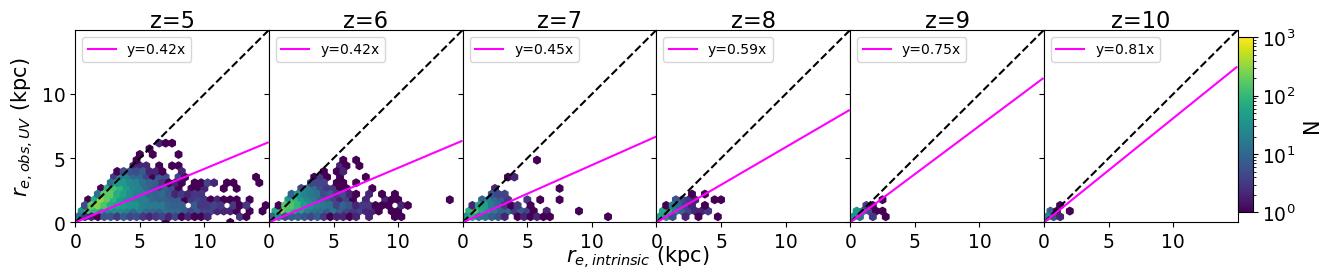}
     \caption{Observed UV sizes from mock images (y-axis) are plotted as a function of intrinsic sizes (x-axis) at SNR=5. The black dashed line is a 1:1 relation whereas the magenta line is the best fit y=mx relation where y is observed size and x is intrinsic size.}
     \label{fig:sizecomp}
\end{figure*}

\subsubsection{IR imaging}\label{sec.alma}
We create imaging for at high resolution IR for a FIR hypothetical instrument similar to NIRCam. To also compare to current observational studies, we also simulate mock ALMA imaging of the FIR dust-continuum at rest frame 158 \um. We use \texttt{simobserve} task in CASA \citep{CASA},  to create the measurement set. We simulate ALMA observations with two different resolutions. All three of these IR imaging techniques are described below:

%\subsubsection{ALMA Simulations}

\begin{itemize}
    \item \textit{High Resolution NIRCAM like imaging:} We use the per pixel SED data for the far-infrared from our radiative transfer simulations. At these high-redshifts, there are no similar observatories (without unrealistic observational time required) with as high a resolution as \jwst\ in the rest-frame IR (the highest resolution channel on the \textit{Herschel Space Observatory} was 5 arcsec), 
    we create an arbitrary PSF from a 2D Gaussian kernel with a standard deviation of 2 pixels (0.052" at $z=5$) for adding observational effects to our images. 
    Gaussian and shot noise are also added. We simulate noise for SNR= 5, 10, 15 and 20 but stick with using SNR = 5 for all further analysis. We do this process for 50\,\um\ and 250\,\um\ wavelengths. There is no significant sizes evolution with  wavelength for our sample as shown in \S \ref{size evolir}, and hence we use the 50 \um\ mock photometry for all further analysis. This being a simulated photometric imaging, the sizes for these images will be denoted by $R_{e,obs,IR,phot}$.

    \item \textit{ALMA Low angular resolution:} We also simulate observations with similar parameters as presented in the CRISTAL survey \citep{mitsuhashi2023almacristal}. We use configuration C3 for ALMA, which leads to a resolution of $\approx 0.3$". The resultant images  (for the same sample of galaxies as in Figure \ref{fig:NIRCamfalse}) are shown in Figure \ref{fig:ALMAfalse}. The sizes for these images will be denoted by $R_{e,obs,IR,alma-lr}$.

    \item \textit{ALMA High angular resolution}: We simulate observations with $\approx 0.01-0.02$" angular resolution hence requiring the extended C-8 configuration. The resultant images produced are shown in Figure \ref{fig:ALMAfalsehighres}. The sizes for these images will be denoted in further studies by $R_{e,obs,IR,alma-hr}$.
    
\end{itemize}
 For both ALMA categories, observation is done with a bandwidth of 7.5 GHz around the redshifted frequency (rest-frame 158\,\um) for our observations, with SNR=10 level of Gaussian noise induced before ALMA simulation. ALMA beam effect is simulated on this preinduced noisy image. Due to adopting a fixed sky temperatures and zenith opacity, we observe ALMA SNR in the range of [3,15]. For our sample an integration time of approximately 2 hours and 40 hours per object would be required to achieve a SNR of 3 and 15 respectively for the ALMA low-resolution imaging. SNR of most of the sample at z= 5, 6 matches well with the observed SNR in CRISTAL survey \citep{mitsuhashi2023almacristal}. We set a sky temperature of 260K for the simulation with the zenith opacity being 0.1 at the observing frequency. To analyse the simulated observations, we clean the obtained output by running Tclean, the inherent cleaning algorithm in CASA, for 10000 iterations, deconvoluting the data using Hogbom CLEAN algorithm \citep{1974A&AS...15..417H}. The maximum and minimum depth of cleaning are 0.05 and 0.8 times the PSF fraction respectively.

\section{Effect of noise and PSF}\label{sec:noiseeffect}
PSF describes the spread of the flux over a range of pixels for a point source due to the diffraction of light entering the optics of an observatory. The optics spreads the sources in a pattern around its surroundings, making the surrounding regions of bright sources even brighter. 
Along with this, Gaussian noise will increase or decrease the brightness of a pixel due to randomness of noise assignment. 
The relative brightness of very faint/median pixels can change significantly compared to the brightest pixel due to the addition of noise. Since we calculate half-light radius by measuring total luminosity inside a certain aperture and compare it with the total luminosity (hence dealing in relative brightness), this jumping of pixels to different brightness bins affects the measured sizes.

We took a sample of images from different redshifts and normalized them and then calculated average brightness of a pixel at different radii (radii are calculated from the centre of potential defined in the simulations to perform consistent analysis across all galaxies). Figure \ref{fig:brightness comparision10} compares the original radiative transfer output to the noisy image. We measure and compare the shift in luminosity profiles of the intrinsic and observed images to see the effect of noise with increasing redshift. The higher the redshift, the more the median brightness pixels are closer in magnitude to the maximum brightest pixels. It is also apparent that some luminosity of a galaxy will be lost due to noise reduction while applying a minimum threshold pixel value, also discussed in \citet{varadaraj2024sizes}. This leads to fewer pixels being evaluated for size calculation.  We show the mean percentage of pixels for galaxies in each redshift from our sample in bins of relative brightness 
% bin difference for our galaxies from 
for intrinsic and observed images in Tables \ref{tab:RTComp} and \ref{tab:ImageComp} respectively. This impact of noise is dependent on observational parameters and filters used, affecting the depth of observations. \citet{varadaraj2024sizes} finds that low wavelengths filters (F115W) find a larger impact of noise on sizes than higher wavelength filters (F444W). Noise washes out faint sources effectively at low depth leading to higher offset between intrinsic sizes and the observed sizes  at higher depths. We see a similar effect of noise which causes the underestimation of observed sizes. On the other hand, the PSF boosts the size of bright parts of the galaxy. This can be seen in Figure \ref{fig:psfeffect}, where at SNR=15 the sizes of mock images simulated without effect of PSF added on, all observed sizes were underestimated compared to intrinsic sizes with a factor of 0.48-0.53 times. It can be seen later in \S \ref{intrsize} that at the same SNR, observed images show overestimation in sizes for higher redshift.  This comparison also shows that the underestimation of sizes due to noise is consistent with increasing redshift, as noise primarily washes out only the faint regions of the galaxy.  The effect of PSF is more significant for more compact galaxies, as bright sources when spread by the PSF, become closer (see Figure \ref{fig:compact}).

Combining the observational effects of noise, pixel scale and PSF, our study shows that at lower redshifts (z= 5-8) galaxy centres are not extremely bright relative to the median luminosity regions, and hence Gaussian noise and PSF are able to move many pixels to a higher luminosity. Due to the effect of PSF, observed images have bright centres that are spread out, while the low luminous parts of a galaxy blend with noise. The loss of luminosity to noise triumphs the effect of PSF at lower SNR resulting in a increment of the r20 (radius at which 20 percent light is present) and a reduction of r80 (radius at which 80 percent light is present). This leads to underestimation (0.4-0.6 times)  of the galaxy sizes at lower redshifts. With increasing redshift, the galaxy centres are comparatively brighter and compact which is further strengthened by PSF. This leads to underestimation of sizes with higher factors (0.7-0.9 times) compared to lower redshifts. With increasing SNR and redshift, lesser number of low luminous parts of the galaxy blend into the noise. As the \flares\ galaxies are more compact at higher redshift, the loss of low luminous regions is compensated by the PSF in bright centres, leading to similar r20. The luminosity shift to the centre remains larger than the luminosity lost due to noise, hence r80 and r50 increases. Thus, the sizes of galaxies are being overestimated at high redshift and high SNR. This change of r20 and r80 can be seen in Figure \ref{fig:brightness comparision10} for individual samples at z= 5, 7, 10 for which the sizes are underestimated, nearly same, and overestimated respectively. Figure \ref{fig:r20r80} also shows the change in r20 and r80 described above, averaged over the whole sample for galaxies with underestimated and overestimated sizes.

\begin{table}
    \centering
    \begin{tabular}{ccccccc}
    \hline
        \textbf{Relative} & \multicolumn{6}{c}{\textbf{z}} \\
        \textbf{Brightness} & \textbf{5}& \textbf{6} & \textbf{7} & \textbf{8} & \textbf{9} & \textbf{10}\\ \hline \hline
        (0.0,0.2] & 96.49& 95.33& 93.81& 91.23& 87.63& 85.45\\ 
        (0.2,0.4] & 2.60&2.87&3.53&4.43&5.59&76.24 \\ 
        (0.4,0.6] & 0.66&1.16&1.58&2.33&3.64&4.14 \\ 
        (0.6,0.8] & 0.18&0.46&0.71&1.31&1.78&2.17\\
        (0.8,1.0] &  0.06&0.18&0.36&0.70&1.35&2.00 \\ \hline
    \end{tabular}
    \caption{Mean percentage of Pixels is shown in each relative brightness bins after normalizing the intrinsic images in a square aperture of width equal to 4$R_e$ ($R_e$ is intrinsic RT size).}\label{tab:RTComp}
\index{tables}
\end{table}

\begin{table}
    \centering
    \begin{tabular}{ccccccc}
    \hline
        \textbf{Relative} & \multicolumn{6}{c}{\textbf{z}} \\
        \textbf{Brightness} & \textbf{5}& \textbf{6} & \textbf{7} & \textbf{8} & \textbf{9} & \textbf{10}\\ \hline \hline
        (0.0,0.2] & 0.40& 0.41& 0.19& 0.25& 0.13& 0.06\\ 
        (0.2,0.4] & 29.18&25.85&18.64&22.56&12.48&10.02 \\ 
        (0.4,0.6] & 59.28&61.23&61.95&56.70&48.37&44.97 \\ 
        (0.6,0.8] & 10.02&10.93&15.76&15.74&28.82&35.99\\
        (0.8,1.0] & 1.11&1.58&3.47&4.76&10.21&8.96 \\ \hline
    \end{tabular}
    \caption{Mean percentage of Pixels is shown in each relative brightness bins after normalizing the observed (SNR=5) images in a square aperture of width equal to 4$R_e$ ($R_e$ is intrinsic RT size).}\label{tab:ImageComp}
\index{tables}
\end{table}
\section{UV size analysis} \label{sec:UVsizes}

In this section we analyse how the UV sizes evolve with cosmic time, luminosity, mass and also wavelength. We compare sizes in our study with prominent observational and simulation studies. We also analyse the variation of galaxy sizes at 1500 \AA\  between simulations and mock observations.

\begin{figure*}
\centering
   \includegraphics[width=17.5cm]{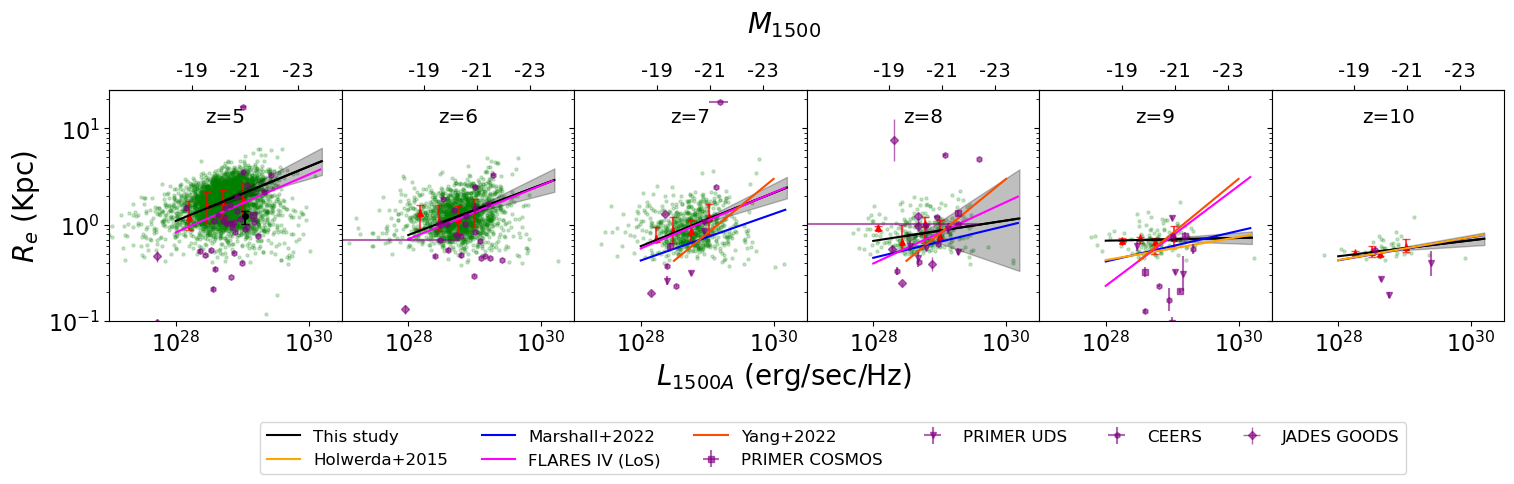}
     \caption{The UV luminosity size relation at different redshifts is shown. The y-axis shows the UV 1500 \AA\ sizes in kpc and the x-axis shows luminosity at the same wavelength. The green scatter shows our sample of galaxies analysed with radiative transfer and then simulated observations for NIRCAM. The red plots show the median luminosity and sizes in bins of 0.3dex in luminosity with 16th and 84th percentile errorbars. The black line is an linear fit to the median luminosity and sizes fit to the power law.  We also compare our sample with observational studies \citep{Holwerda_2015,Bouwens_2022} and simulations \citep[][\citetalias{10.1093/mnras/stac1368}]{10.1093/mnras/stac380}. \jwst\ data has been taken from \citet{morishita2023enhancedsubkpcscalestarformation} and \citet{2024MNRAS.527.6110O} for the sample with $\text{M}^*/\textup{M}_\odot \geq 10^9$ from JADES \citep{eisenstein2023overviewjwstadvanceddeep}, CEERS \citep{2024MNRAS.529.1067H} and PRIMER survey.}
     \label{fig:lumsize7}
\end{figure*}

\subsection{Intrinsic vs observed sizes} \label{intrsize}
Mock observations using simulations have shown stark difference between observed and intrinsic size evolution relations. Galaxies in the \textsc{bluetides} simulation \citep{10.1093/mnras/stac380} show a negative dust-free size luminosity relation at z=7, which when evaluated for dust attenuation results in a positive relation. Similarly in \flares\ (\citetalias[][]{10.1093/mnras/stac1368}) negative dust-free size stellar mass and  dust-free size luminosity relations also show a positive correlation for dust attenuated sizes. Although the \textsc{Simba} simulations \cite[]{10.1093/mnras/stz937}, uniquely shows positive correlation for  dust-free and observed size with mass \citep{10.1093/mnras/staa1044}. \citet{2023ApJ...946...71C} also found simulated observed sizes for massive galaxies at 2 \um\ and 3.6 \um\ to be larger than  dust-free stellar mass sizes at lower redshift (z= 3, 4), attributing it to mass being more compact than observable stellar light, predicting heavy dust obscuration.

Similar to the above studies, we compare variation between mock observational sizes and our intrinsic sizes to analyse the observational effects causing the variation in size evolution relations. Figure \ref{fig:sizecomp} compares the sizes from mock observations and intrinsic sizes for each of the redshift (0.40-1.16 times) at SNR=5. The best fit size relation at other SNR=10, 15 and 20 is also presented in Table \ref{tab:compslope}. The figure shows a significant variation in observations and simulation sizes. From Table \ref{tab:compslope} it can be clearly seen that observational sizes will be underestimated (0.40-0.94 times) for all redshifts for SNR<=10. For higher SNRs for lower redshift ($z \in [5,6,7,8])$ sizes will be underestimated (0.64-0.83 times) while the sizes for higher redshift ($z \in [9,10]$) will be slightly overestimated (1.04-1.16 times) compared to intrinsic sizes. \thesan\ simulations \citep{2024arXiv240208717S} also shows a similar variation in size with galaxies with $10^7 >$ $\text{M}_*/\textup{M}_\odot$> $10^9$  having median simulation size being greater than up to 3 times the observed sizes with increasing stellar mass. This variation in sizes is primarily due to observational effects (PSF and noise) as the offset is independent of galaxy properties like stellar mass and FUV luminosity (see Figure \ref{fig:sizedep1}). Further explanation of observational effects is present in \S \ref{sec:noiseeffect}.

Sizes from mock images are calculated by \texttt{statmorph}, while as centre of potential is known to us from simulation,  we calculate  the intrinsic sizes using the Iterative Aperture Method (as discussed in \S \ref{IterAp}). With the effect of noise decreasing with increasing SNR at high-redshift we see that most sizes are overestimated. Noise being primarily responsible for washing away fainter parts of the galaxy along with the PSF intensifying the bright parts of the galaxy, this effect with increasing SNR shows the difficulty accounting for the complex PSF which causes overestimated sizes (see \S \ref{sec:noiseeffect}).

\begin{table}
    \centering
    \begin{tabular}{c|cccc}
    \hline
        \textbf{Redshift(z)} & \multicolumn{4}{c}{\textbf{Slope of the Best Fit Line}} \\
        & \textbf{SNR 5}& \textbf{SNR 10}& \textbf{SNR 15}& \textbf{SNR 20}\\ \hline \hline
        5 &  0.42 & 0.54 & 0.64 & 0.70\\ 
        6 &  0.42 & 0.56 & 0.64 & 0.73\\ 
        7 &  0.45 & 0.56 & 0.63 & 0.71\\ 
        8 &  0.59 & 0.71 & 0.83 & 0.89\\
        9 &  0.75 & 0.91 & 1.04 & 1.10\\
        10 & 0.81 & 0.94 & 1.07 & 1.16\\ \hline
    \end{tabular}
    \caption{Slopes of Best Fit Line (y=mx, no intercept) where y is mock image effective radius and x is simulation effective radius at different signal to noise ratios }\label{tab:compslope}
\end{table}
\begin{table}
    \centering
    \begin{tabular}{ccc}
    \hline
    \textbf{Redshift} & \textbf{$\beta$ for Intrinsic Sizes} &  \textbf{$\beta$ for Observed Sizes} \\ \hline \hline
       5 & 0.4196 $\pm$  0.0121 & 0.2820 $\pm$ 0.0821\\ 
        6 & 0.4850 $\pm$ 0.1000 & 0.2595 $\pm$ 0.0816\\ 
        7 &  0.5971 $\pm$ 0.1387 & 0.2769 $\pm$ 0.0701\\ 
        8 & 0.2808 $\pm$ 0.3568 & 0.1058 $\pm$ 0.3446\\
        9 &  0.1499$\pm$ 0.2365 & 0.0141 $\pm$ 0.0395\\
        10 & 0.2456 $\pm$ 0.1240 & 0.0829 $\pm$  0.0392\\\hline
    \end{tabular}
    \caption{$\beta$ at different redshifts for radiative transfer output and the mock images (SNR=15) created.}\label{tab:BetaSlopes}
\end{table}
\subsection{UV size-luminosity relation} \label{sec:sizelum}

\begin{figure*}
\centering
   \includegraphics[width=16cm]{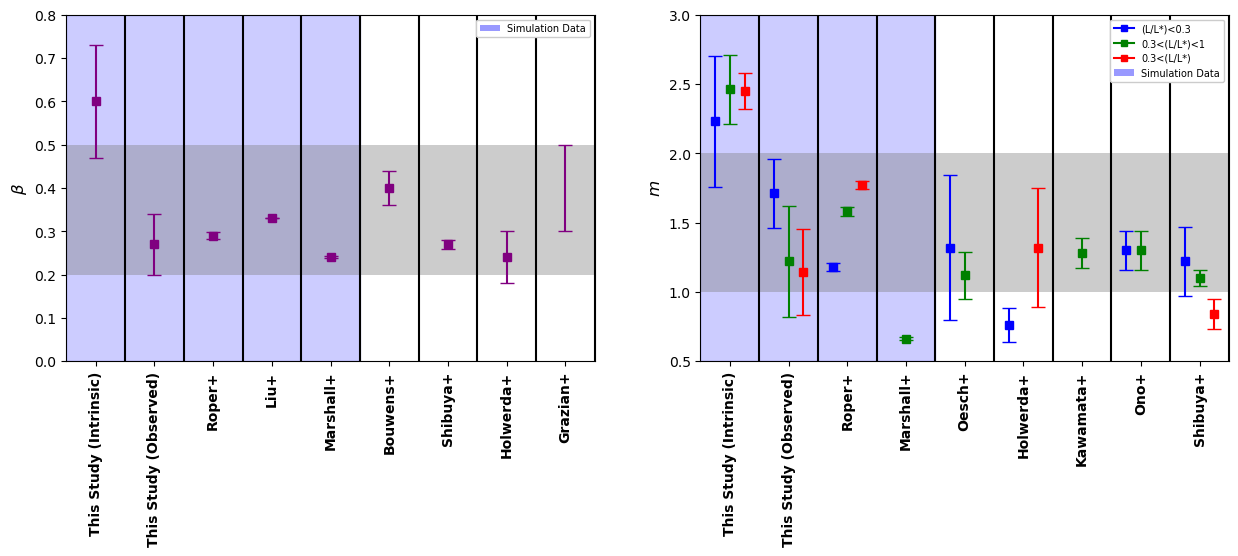}

     \caption{Left: Comparison of the size-luminosity slopes ($\beta$) in relation with other studies \citep{10.1093/mnras/stac1368,10.1093/mnras/stw2912,10.1093/mnras/stac380,Bouwens_2022,Shibuya_2015,Holwerda_2015,Grazian} at z=7 (blue shaded region are simulation studies) is shown. Right: Comparison of the UV size evolution slopes (m) in relation with other studies \citep{10.1093/mnras/stac1368,10.1093/mnras/stac380,Oesch_2010,Holwerda_2015,Kawamata_2018,Ono_2013,Shibuya_2015} is shown.}
     \label{fig:sizeevol}
     \label{fig:lumcomp}
\end{figure*}

We binned our data in luminosity bins of 0.3 dex and calculated the median luminosity and median sizes in these bins. We fit Equation \ref{eqn:lumpowerlaw} to these medians. Figure \ref{fig:lumsize7} shows a luminosity-size fit for our sample across different redshifts compared to observational and simulated data.

 %It is also worth noting that, our sample lies higher than intercepts of other studies due to a bias towards more massive objects.
 We see a positive $\beta$ in agreement with major observational studies \citep{Holwerda_2015,Grazian,Bouwens_2022}, with our observed size slopes matching these studies well at $z=5,6$ and $7$ but the slopes at $z>7$ are shallower with high-error margin. Our intrinsic slopes are significantly higher ($>\times2$) than our observed slopes. Observations compared to simulations lack very dispersed galaxies in their sample. For our dispersed galaxies at lower SNR (<10), with low surface brightness, majority of the galaxy, blends with noise, leading to very compact observed sizes. These dispersed galaxies also flatten the size luminosity fit, leading to lower observed slope values than our intrinsic slopes. Also in contrast to other studies we see a non-evolving  $\beta$ \cite[see Table \ref{tab:BetaSlopes}, similar to][]{Shibuya_2015} for $z=5,6$ and $7$. In Figure \ref{fig:lumcomp} we compare our $\beta$ slopes for both intrinsic and observed sizes with other simulations and observations at $z=7$, a common redshift that is used in comparisons across simulations and observations at high-redshift. When compared with observations \citep{Holwerda_2015, Huang_2013, Bouwens_2022}, our simulated observed slope at SNR=5 is in agreement with all the studies. Our observed slopes are a bit shallower than early JWST studies like \citep{Yang22} but compared to recent observations \cite[CEERS, PRIMER \& JADES data taken from][]{morishita2023enhancedsubkpcscalestarformation,2024MNRAS.527.6110O} we match the scatter at z $\geq$ 6 well. Although we do find a lack of ultra compact, low luminosity sample in our dataset at z= 5, 6.

We also plotted size-luminosity relations at different wavelengths to see how the effect of dust evolves as a function of wavelength (see Figure 
\ref{fig:dusteffect}). When the effect of dust is evaluated at a fixed redshift, dust causes the total luminosity to decrease and as a result size increases due to dust obscuring the bright centres of the galaxies. With increasing wavelength, the decrease in luminosity is lesser, while the sizes decrease. This makes the best size-luminosity fit decrease with increasing wavelength in the Figure. This follows well with the work in \citet{10.1093/mnras/stac380} and \citetalias[][]{10.1093/mnras/stac1368}.

 A comparative luminosity-size relation with non-parametric size calculation is present in Figure \ref{fig:lscm}

\subsection{UV size evolution with redshift}
We also explore the evolution of sizes with redshift. Since intrinsic luminosity is related to stellar mass, we bin the data in mass bins of 0.5 dex. Higher mass galaxies tend to have a higher luminosity. We then choose the mass bin of $9.5 < M^*/\textup{M}_\odot\ <10$ which is commonly used in literature \citep{Shibuya_2015,Mosleh_2012}. This bin is also sufficiently populated in all redshifts with a good spread of luminosities. As the sizes vary with luminosities as shown previously, we fit Equation \ref{eqn:Power Law}, for our sizes, for this mass bin as three different luminosity samples:
\begin{enumerate}
    \item $0.3 > \frac{\text{L}_{UV}}{\text{L}^*_{z=3}}$
    \item $ \frac{\text{L}_{UV}}{\text{L}^*_{z=3}} > 0.3$
    \item $1 > \frac{\text{L}_{UV}}{\text{L}^*_{z=3}} > 0.3$
\end{enumerate}

For the UV luminosity bins, the observed size slopes are m = $1.71{\pm 0.25}$, $1.14{\pm 0.31}$, $1.22{\pm 0.40}$ and  $R_0$ = $30.10 \pm 14.67$, $13.12 \pm 10.30$ and $11.61 \pm 6.96$ respectively\footnote{We use curve\_fit (non-linear least-squares fitting), from \texttt{scipy} \citep{Virtanen_2020}, to produce fits and the error are generated by the covariance matrix.}. Figure \ref{fig:sizeevol} compares our slopes with other previous \flares\ work and simulation studies and shows higher slopes for intrinsic sizes, with m = $2.23{\pm 0.47}$, $2.46{\pm 0.25}$, $2.45{\pm 0.13}$ and $R_0$ = $112.73 \pm 101.59$, $205.68 \pm 99.40$ and $197.23 \pm 48.73$. This difference in slope evolution is largely due to the observational effects described in \S \ref{sec:noiseeffect} and \ref{intrsize}, with the underestimation of observed sizes decreasing with increasing redshift, the size evolution curve becomes flat with high intercept ($\text{R}_0$) factors. Intrinsic sizes for all three luminosity bins have similar median sizes at z=5 but with increasing redshift lower luminosity sample tends to be more compact than higher luminosity sample. This results in low luminosity bins having lower intercepts and slopes, which also follows well with previous \flares\ work \citep{10.1093/mnras/stac1368}. For observational sizes these lower luminosity sample being more compact leads to a stronger effect of PSF flattening the evolution curve, which leads to higher slopes and intercepts with m=1.71 for the lower luminosity bin and m=1.14 and 1.22 for higher luminosity bins due to fit being flattened. Figure \ref{fig:sizeevol} also compares our intrinsic and observed size fits to other studies. For observed sizes, slopes of all our luminosity bins fall within the range  m=$1.0-2.0$, similar to \citet{2018MNRAS.477..219M} and also near the viral radius size-evolution slope m=1.0 and m=1.5 of dark matter halos \citep{1998MNRAS.295..319M,2002MNRAS.336..112M,2004ApJ...600L.107F}. We are also in agreement with \citet{Oesch_2010} for $0.3\text{L}^*_{z=3} > {\text{L}_{UV}}{}$ and $\text{L}^*_{z=3} > {\text{L}_{UV}}{} > 0.3\text{L}^*_{z=3}$ bins, while also matching \citet{Shibuya_2015,Ono_2013,Kawamata_2018} for the later. \citet{Holwerda_2015} is the one of the very few studies for very bright galaxies at high redshift stating a m=$1.32 \pm 0.43$ and our brightest luminosity bin matches within the error margin of that study.

\subsection{UV size evolution with stellar mass} \label{size evol}
UV size evolution against stellar mass has been analysed in simulations using \thesan\ \citep{2024arXiv240208717S}. The dust-free sizes of galaxies show a negative correlation with increasing stellar mass ($ \text{M}^*/\textup{M}_\odot \geq 10^{8.0}$) with the dust-attenuated sizes showing a positive correlation with increasing stellar mass. %\flares\ half-mass sizes  also show a bi-modal negative size-stellar mass correlation \citep{roper2023flares}, indicating two different tracks of galaxy growth. 
Our intrinsic sizes (noise and PSF free) with \flares\ also shows a positive correlation with stellar mass as our sizes also account for radiative transfer effects of dust attenuation and extinction. \citetalias[][]{10.1093/mnras/stac1368} has also shown that dust attenuated sizes can be nearly 50 times the simulation sizes. 

\begin{figure}[]
	\centering
		\includegraphics[width=7cm,keepaspectratio]{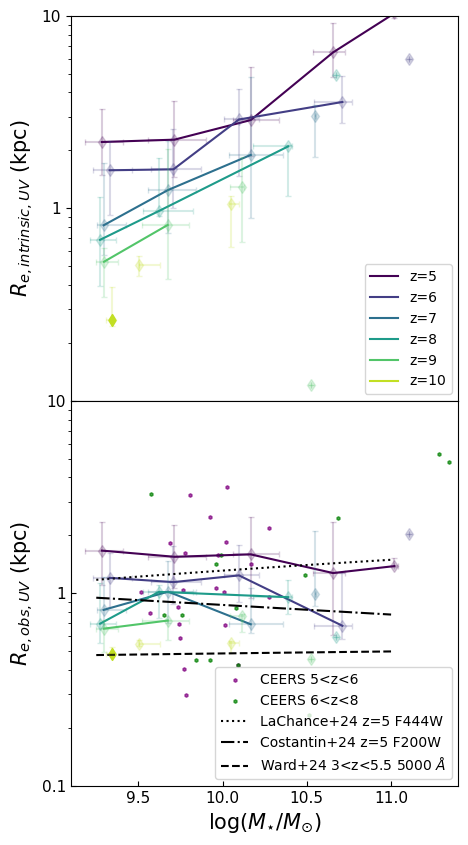}
	\caption{Upper panel: The evolution of the UV simulation sizes (y-axis, in kpc) as function of stellar mass (x-axis) for $z\in[5,10]$. Lower panel: Evolution of sizes from mock observations as function of stellar mass. Size mass relation for simulated observations from \citet{2023ApJ...946...71C,2024arXiv240116608L} and NIRCAM observations from \citet{ward2023evolutionsizemassrelationstarforming}. CEERS data has been taken from \citet{morishita2023enhancedsubkpcscalestarformation,2024MNRAS.527.6110O}.}
	\label{fig:masssizeevol}
\end{figure}

Figure \ref{fig:masssizeevol} shows median galaxy stellar masses (bins of $10^{0.5} \textup{M}_\odot$) plotted against the intrinsic and observed UV sizes. The intrinsic UV sizes of galaxies show an increasing slope with increasing stellar mass. For galaxies with the same stellar mass, those at lower redshift will be larger in size. Adding observational effects, we find a near flat evolution at all redshifts for observed sizes which turns negative at higher stellar mass ($ \text{M}^*/\textup{M}_\odot \geq 10^{10.5}$). The effect of noise in observations eats away at the faint galaxy regions outside bright cores reducing the size for galaxies with non-compact complex morphologies. This effect gains significance the more massive a galaxy is, flattening the observed evolution curve compared to the intrinsic one. This is further supported by the fact that the evolutionary trends of intrinsic size are successfully recovered for mock observations with minimal noise (SNR=50 and SNR=100 size-mass relations are shown in Figure \ref{fig:masssnr100}). Similar to our study \cite{2023ApJ...946...71C}, using \mbox{\sc Tng50}, simulated NIRCam observations using radiative transfer. They also predict near flat observed size-stellar mass relation at $z=5$ which turns negative at $z=6$. \citet{2024arXiv240116608L} using \mbox{\sc Astrid} simulations, finds flat slopes for simulated NIRCam F444W observations at $z= 5, 6$ while also finding the correlation between mass and size decreasing with increasing redshift (from $z=3$ to $z=6$). As seen in Figure \ref{fig:masssizeevol} our observed sizes match the observational scatter of CEERS z=5-8 galaxies from \citet{2024MNRAS.527.6110O}. Observations compared to simulations, particularly \flares, lack statistically significant number of galaxies at higher mass end ($ \text{M}^*/\textup{M}_\odot \geq 10^{10.5}$), making the high redshift size-mass relation difficult to constraint. Although in recent CEERS survey, \citet{ward2023evolutionsizemassrelationstarforming} has shown a positive correlation between stellar mass and sizes (at 5000 \AA). Compared to \citet{ward2023evolutionsizemassrelationstarforming,2023ApJ...946...71C,2024arXiv240116608L} we have higher intercepts due to probing observed sizes at a shorter wavelength.

\section{IR size analysis} \label{sec:IRsizes}

We compare sizes in FIR regime by various methods used to study the spectrum. We also discuss the size evolution against cosmic time and stellar mass while also drawing comparisons to observational studies.

\subsection{Comparison of IR sizes by various methods}
\begin{figure}
    \centering
   \includegraphics[width=8cm]{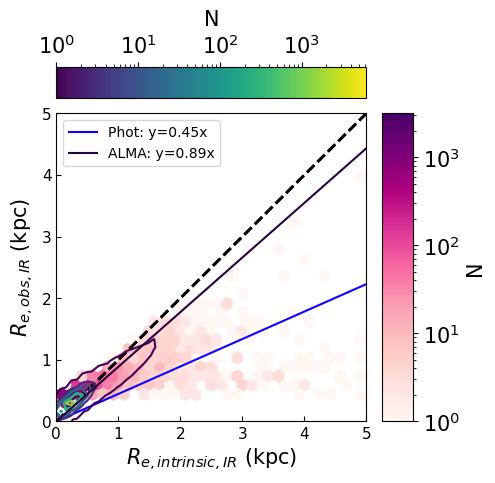}
     \caption{Comparison of sizes between various IR observation methods is shown above. The plot compares IR sizes (y-axis) from mock photometry (50 \um) and ALMA simulations (high-resolution, $\approx 0.01-0.02''$ at 156 \um) with intrinsic sizes (50 \um) (x-axis). The hex-bin plot shows the log mapped distribution of photometric sizes whereas the contours show the log mapped distribution of mock ALMA sizes. The black line shows the 1:1 relation with the yellow line showing the best-fit y=mx relation. All ALMA sizes in this plot of high angular resolution ($\approx 0.01''$ angular resolution)}
     \label{fig:ircomp}     
\end{figure}

We use the IR photometric images and ALMA images produced as described in \S \ref{sec.alma} and calculate the sizes using using \texttt{statmorph} as discussed earlier. Intrinsic sizes are bigger than both high-resolution ALMA and photometric sizes. Figure \ref{fig:ircomp} compares the intrinsic sizes measured using the high-resolution observation methods, 50 \um\ photometry and ALMA mock imaging ($\approx 0.01-0.02'' $ angular resolution). It shows that interferometric ALMA sizes at high resolution will be able to measure sizes with a higher accuracy than NIRCam like photometry in IR when evaluated for SNR $\approx$ 5. This is due low luminous emissions blending with noise in photometry. For ALMA at this high resolution, the low-surface brightness extended emissions of a galaxy will be lost. The effect of noise and PSF, similar to UV, are present in photometric imaging. This can be seen when comparing Figures \ref{fig:ALMAfalsehighres} \& \ref{fig:realIR} at $z=6$ sample's lowest mass bin. 

\begin{figure}
    \centering
   \includegraphics[width=8cm]{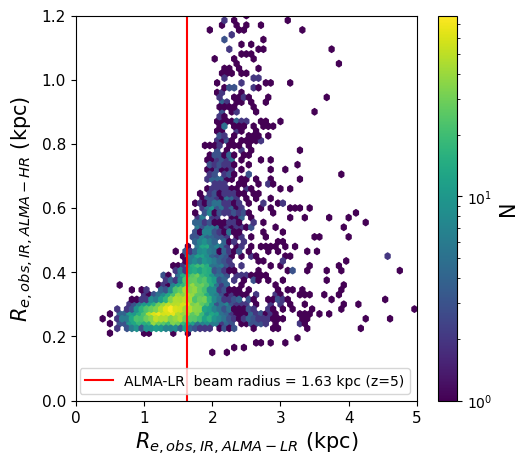}
     \caption{Comparison of 158\um\ sizes between low ($\approx 0.3''$) and high ($\approx 0.01''$) resolution ALMA configuration setup is shown above for z=5, 6 sample. The high resolution ($\approx 0.01''$) sizes is on y-axis with  low resolution ($\approx 0.3''$) sizes on x-axis. The line shows the average beam radius for the low-resolution observations.}
     \label{fig:c3c10}
     
\end{figure}

For ALMA observations at lower resolutions than this ($\approx 0.3'' $ angular resolution), small scale structures in the galaxy can be resolved to a certain extent. We compared our  $\approx 0.3'' $ lower angular resolution ALMA sizes to the high angular resolution ALMA sizes in Figure \ref{fig:c3c10}. The Figure shows that at $z=5-6$, for ALMA low angular resolution ($\approx 0.3''$) sizes are more than 5 times the high angular resolution sizes. This difference of sizes can be attributed to effect of the larger ALMA beam, spreading the observations of the compact dust core to larger sizes. From comparing Figures \ref{fig:ALMAfalse} \& \ref{fig:realIR}, it can also be seen that at very low SNR, noise close to the IR emission can act as extended observation, leading to further increase in sizes (Figure \ref{fig:snreffect} affirms this by showing that at low SNR, the low angular resolution sizes to intrinsic size ratio tends to be higher than at high SNR). Figure \ref{fig:c3c10} also shows that nearly half of the galaxies are unresolved, falling below the beam size (2720 of 5616 or 48.43\% unresolved).

We also compared the ALMA low angular resolution ($\approx 0.3''$) sizes to the recent CRISTAL survey \citep{mitsuhashi2023almacristal,ikeda2024almacristalsurveyspatialextent}, observed at similar resolutions. We find that the dust continuum sizes around [C\textsc{ii}] emission in CRISTAL galaxies matches well with our dust-continuum sizes from mock ALMA observations at same wavelengths and resolutions. CRISTAL sizes range from 0.6-2.44 kpc with a mean of 1.38 kpc at z=4-6 whereas for similar redshift range mock ALMA sizes for \flares\ range from 0.42-3.32 kpc (99th percentile). With the intrinsic sizes of galaxies in \flares\ for the same mass and redshift range as CRISTAL galaxies being much lower ($\approx10$ times) than the beam size of observations while the observed sizes being comparable, alludes to the beam effects enlarging the core emission of galaxies similar to the effect of PSF leading to inflated ALMA sizes at low angular resolutions.  %From Figures \ref{fig:ALMAfalse} \& \ref{fig:realIR}, it can be seen that the combined effect of the beam size on observations as well as very low SNR close to the IR emission can act as extended emission, leading to further higher sizes (See Figure \ref{fig:snreffect}.

% Along with the effect of beam on observations, as seen by comparing  at very low S/N noise close to the IR emission can act as extended emission, leading to higher sizes (See Figure \ref{fig:snreffect}).  
%\end{figure}

\begin{figure}
    \centering
   \includegraphics[width=7.5cm]{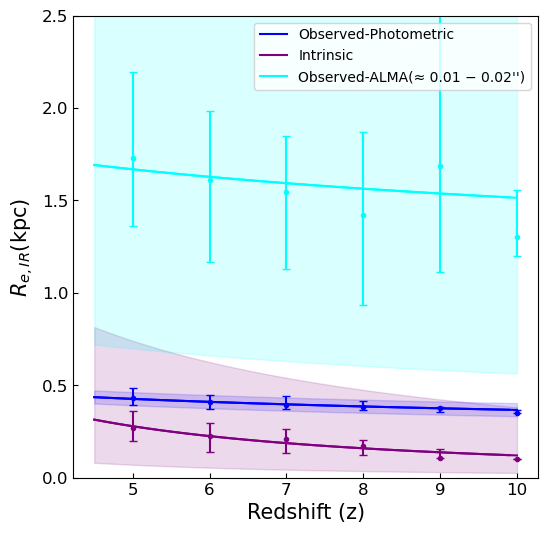}
     \caption{Infrared Size Evolution with redshift at 50 um rest is shown with the half-light radius for both the simulations and mock-photometry on the y-axis and the redshift on the x-axis. The shaded areas highlight the error margins for the fit.}
     \label{fig:irsizeevol}
     
\end{figure}
\begin{figure}[]
	\centering
		\includegraphics[width=7.5cm,keepaspectratio]{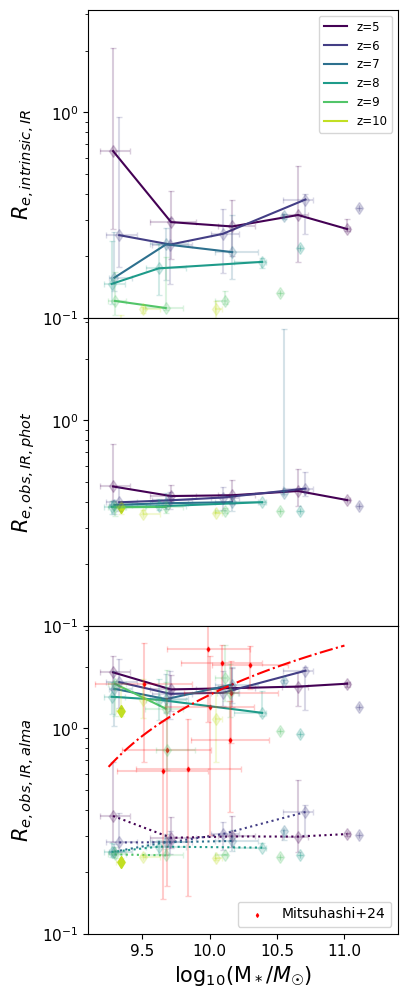}
	\caption{Upper panel: The median intrinsic sizes(y-axis) of 50\um\ in simulations as a function of stellar mass(x-axis) are shown.  Middle panel: The evolution of mock IR photometric observational sizes at SNR=5. Lower panel: The evolution of ALMA observed sizes. The ($\approx 0.01-0.02''$) angular resolution sizes are represented by the dotted lines and the low angular resolution ($\approx 0.3''$) sizes are represented by the solid lines. The red scatter and the corresponding line show the CRISTAL \citet{mitsuhashi2023almacristal} data and its best fit mass-size relation respectively. }
	\label{fig:masssizeevolir}
\end{figure}

\subsection{IR size evolution with redshift}
We evaluated the evolution of sizes in the infrared with the same process used for UV. We calculate intrinsic sizes with the iterative aperture method and also use \texttt{statmorph} to calculate observed sizes for the photometric and ALMA imaging. Figure \ref{fig:irsizeevol} shows the evolution of intrinsic and observed IR sizes, which follows a power law when the median sizes are plotted against redshift. There is negligible size evolution at IR wavelengths which leads to shallow slopes, with m=0.24 for photometry, m=0.16 for ALMA at low resolution  and m=1.39 for simulation size. \flares\ galaxy cores were found to be heavily dust attenuated when radial dust attenuation was analysed in \citet{Vijayan2024}. Our shallow slope values with intrinsic sizes being $r \leq 1$ kpc also indicates that majority of the dust produced in the early universe is concentrated in compact cores. The observed IR size-evolution (Figure \ref{fig:irsizeevol}) is also in agreement with ALMA observations of sub-millimetre galaxies \cite[SMGs,][]{Gullberg19}, showing a weak decline of dust-continuum sizes with increasing redshifts (up to $z=5.5$).  Further analysis with respect to comparison of sizes in IR against UV is discussed further in \S \ref{sec.uvir}.

\subsection{IR size evolution with stellar mass} \label{size evolir}
To analyse the IR size evolution with stellar mass, we study their median evolution in bins of 0.5 dex stellar mass across each redshift. Figure \ref{fig:masssizeevolir} shows the evolution of intrinsic IR sizes with stellar mass. At $\text{M}^*\geq 10^{10.0} \textup{M}_\odot$, IR sizes exhibit a slight positive correlation with increasing mass. Figures \ref{fig:masssizeevolir} also shows the IR size evolution against stellar mass for photometric observations and ALMA low/high resolution observations respectively. The photometric sizes of galaxies show no significant evolution with increasing stellar mass and similar to UV, for same stellar mass, galaxies at lower redshift are larger in size. The ALMA low resolution sizes at fixed stellar mass are >10 times the intrinsic sizes. It is important to note that at low ALMA resolution, a lot of the sample is unresolved.  The median ALMA simulated sizes at $\approx 0.3''$ resolution lie well within the scatter of the CRISTAL data \citep{mitsuhashi2023almacristal}. Compared to our simulated ALMA sizes, the best fit size-mass relation for CRISTAL shows a steeper positive evolution of size with increasing stellar mass but more observational data will be required at both the lower ($\text{M}^*\leq 10^{9.5} \textup{M}_\odot$) and higher ($\text{M}^*\geq 10^{10.5} \textup{M}_\odot$) mass end to get a robust infrared size-mass relation.

\begin{figure}
\centering
   \includegraphics[width=7cm]{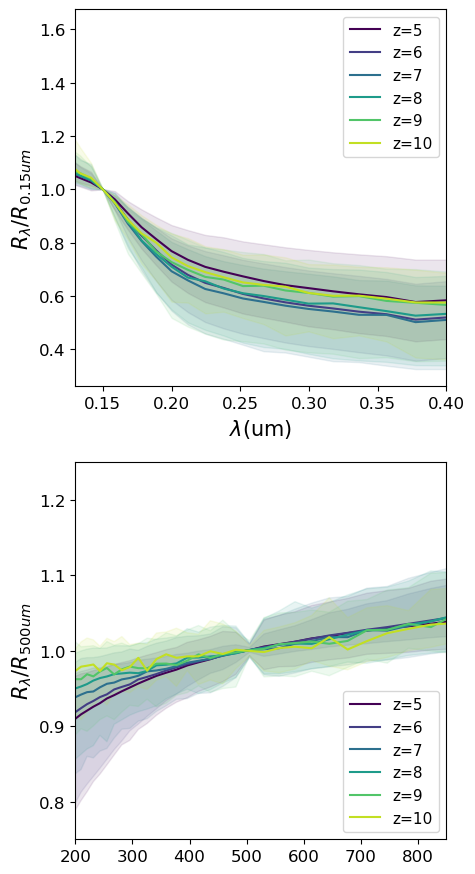}
     \caption{Evolution of normalized size with rest-frame wavelength is shown in UV (upper panel) and IR (lower panel). The solid curves correspond to the median of the
distribution, whereas the colour-shaded regions mark the one-sigma scatter of the distribution}
     \label{fig:uvsizeevolwav}
     
\end{figure}

\section{Panchromatic analysis}\label{sec:pan_analysis}
\subsection{Size evolution with wavelengths}\label{sec:wavevol}

\begin{figure}
\centering
   \includegraphics[width=7.7cm]{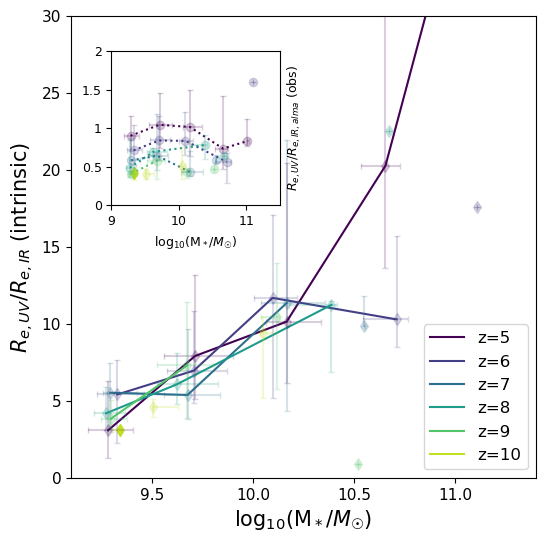}
     \caption{Ratio of intrinsic sizes (main figure) as well mock observations (inset figure) at 1500 \AA\ with observations like ALMA configuration is shown for UV and IR spectrum in stellar mass bins of 0.5 dex, with median values plotted as data-points. The error-bars denote the 16th and 84th percentile values. Only sufficiently populated bins (N>20) have been used to plot the lines.}
     \label{fig:irsizeevolratio}
     
\end{figure}

To evaluate the size evolution with wavelength we first normalise the evolution by dividing the sizes by 1500 \AA\ sizes for UV and 500 \um\ sizes for IR (intrinsic sizes).
Figure \ref{fig:uvsizeevolwav} shows the evolution of median normalized sizes with increasing wavelengths. For each redshift, we observe a decrease (0.4 times the FUV size at 0.4\,\um) in size with increasing wavelength. The size reduction is due to decreasing effect of dust attenuation with increasing wavelengths. At higher wavelengths weak attenuation results in bright centres which capture more luminosity in smaller apertures, reducing sizes effectively. This effect of dust attenuation decreasing with increasing wavelength has been analysed well in \citet{10.1093/mnras/stac380}, \citetalias[][]{10.1093/mnras/stac1368} and our analysis in Appendix \ref{sec.dust}.

Figure \ref{fig:uvsizeevolwav} also shows the evolution of median sizes at FIR wavelengths. These sizes are normalized with the sizes at 500 \um. For FIR, which is not as affected by dust attenuation, we only see a slight positive trend in sizes with increasing wavelengths, which is due to the prevalence of hotter dust (probed by shorter wavelengths) in the centres. \citet{Popping_2021}, using the \tng\ 50 simulation, also showed a similar non-evolution of sizes (normalised at 850 \um\ in the observed-frame) with increasing wavelength in the IR.

\subsection{Comparison of sizes in UV and IR}\label{sec.uvir}
UV emission is pre-dominantly from young stars within galaxies, hence size evolution in UV spectrum is indicative of the progression of unobscured star formation. IR emission helps us to understand the distribution of dust, and compared the UV sizes, IR sizes show shallow evolution either over time or with increasing stellar mass (See \S \ref{size evol} and \ref{size evolir}). This is consistent with studies like \citet{Popping_2021}. The increasing ratio of UV/IR size with increasing mass as seen in Figure \ref{fig:irsizeevolratio} is driven by inside-out growth of star formation, rather than an increase in the dust obscuration within the centres. To analyse this size evolution ratio in UV and IR, we also look at previous \flares\ work to understand the evolution  of star formation and the processes behind it.

\citet[hereafter \flares\ IX]{roper2023flares} details the different
physical mechanisms driving the formation and evolution of compact galaxies. The study presents the physics behind the formation of compact star forming cores in galaxies in the early universe. The star formation criteria in the \flare\ (or \eagle) simulation, imposes a critical gas density \citep{Schaye2004}, which is inversely related to the gas phase metallicity, shown in Equation \ref{eq:sfthreshz} below:
\begin{equation} 
n_{\rm H}^\star(Z)=\min \left [ 0.1\, \cmcubed \left ({\frac{Z}{0.002}}\right)^{-0.64}, 10\,\cmcubed \right]
\label{eq:sfthreshz} 
\end{equation}  \
where $n_{\rm H}^\star(Z)$ is the density threshold and $Z$ is the gas metallicity.\\
% The cores of early galaxies are enriched by metals leading to dense star-forming as well as dust attenuated cores. 
The \flares\ galaxies in the early Universe are metal poor  \footnote{\citep{flares7} shows that at $z=5$, $\textup{Z}^*_{\textup{Fe}}$ for \flares\ is well within the scatter and error-margin of recent observational estimates \citep{Cullen,calabro}. While the sample are still being built at $z>10$, \citet{flares5} states median gas phase metallicity of $\log_{10}(\textup{Z}_g)$ $\approx$ -3  at $ \text{M}^*/ \textup{M}_\odot \approx 10^{8}$, which increases to $\approx$ -2.5 for galaxies with $ \text{M}^*/ \textup{M}_\odot \geq 10^{9}$. This matches recent observations \citep{Hsiao,lake,Harikane,Schouws,Carniani}. Future scope of better statistics at high $z$ still remains.}, and thus the gas density required for star-formation in the simulation is very high. However, once star formation starts in these dense cores, they become very quickly enriched with metals, lowering the critical density for star-formation. This triggers a state of runaway star-formation from this dense gas, quickly enriching them with metals and dust. Thus, these dense regions suffer from very high dust attenuation \cite[see Figure 2 in][]{Vijayan2024}.
At later times the outer regions get enriched with metals, star formation is extended to the outer regions, which also leads to larger sizes in the UV. More details of these processes can be found in \citetalias{roper2023flares}.
 
\citetalias{roper2023flares} finds that massive compact galaxies can form following two different evolution paths. It shows that galaxies at $z=5$ can have progenitors which formed at $z>10$ in pristine environments with low metal enrichment. Stars in such progenitors are formed at high densities, and these galaxies remain compact throughout their evolution unaffected by mergers due to how compact they are (Analysed in Appendix A of \citetalias{roper2023flares}). 
% This is the compact formation path.
The other path, shows that galaxies can also transition from being diffused to compact.  These galaxies have partially metal enriched progenitors (at $z<10$) and have diffused star formation at lower gas densities. They become compact at lower redshifts ($z<6$) when efficient cooling of gas happens in a diffused region, reaching high gas densities. This enables highly localised efficient star formation which enriches the gas in the core of the galaxy, leading to runaway star formation in the core. Hence,  at z=5 galaxies with stellar mass in range of $10^{8.8} \textup{M}_\odot \leq \text{M}^*\leq 10^{9.8} \textup{M}_\odot$ show a significant decrease in size compared to z=6 (See Figure 6 in \citetalias{roper2023flares}). Galaxies with stellar masses below this range ($\text{M}^*\leq 10^{8.8} \textup{M}_\odot$) remain diffused clumpy systems and do not undergo a transition in size.

Figure \ref{fig:irsizeevolratio} shows the UV/IR size ratio as a function of redshift for both radiative transfer sizes with no observational effects as well as mock observations for \jwst\ and ALMA. We can see that for intrinsic sizes for the stellar masses in range $10^{9.0} \textup{M}_\odot \leq \text{M}^*\leq 10^{10.0} \textup{M}_\odot$ , there is a sharp dip in ratio for $z=5$ which can be attributed to galaxies in stellar mass range $10^{8.8} \textup{M}_\odot \leq \text{M}^*\leq 10^{9.8} \textup{M}_\odot$ showing a median decrease in size due to transition from diffused to clumpy systems as presented in \citetalias{roper2023flares}. But when the mock observations for ALMA (low angular resolution similar to \citet{mitsuhashi2023almacristal},$\approx 0.3''$) and \jwst\ are used, the ratio are significantly lower, and IR sizes can be larger than UV sizes, similar to what is seen in \citet{2024A&A...686A.187P,2022ApJ...934..144F,mitsuhashi2023almacristal}. UV/IR Size ratio curve for NIRCAM UV sizes and ALMA high angular resolution sizes, matches the intrinsic curve presented in Figure \ref{fig:irsizeevolratio} as with smaller beam, the ALMA sizes are  predicted closer to the intrinsic IR sizes. A further spatial analysis between UV and IR will be presented in a future study.

\section{Summary and conclusions} \label{sec:conc}
In this study, we perform radiative transfer using \skirt\ \citep{CAMPS201520,camps2020skirt} for galaxies from the \flare\ simulations to study size evolution in the UV and the far-IR at the Epoch of Reionisation ($z \in [5,10]$). 
% This is an alternative approach to the Line of Sight method used by many simulation studies including previous \flares\ work \citep{10.1093/mnras/stac1368,10.1093/mnras/stac380}. 
Radiative transfer using Monte Carlo simulation precisely estimates the effect of dust absorption and scattering of radiation to provide a more accurate Spectral Energy Distribution. We use the results of the radiative transfer simulations to mock observations of these galaxies, for NIRCam at rest frame FUV, to find and analyse the offset in observational sizes to intrinsic sizes of galaxies. We also simulate imaging in the far-IR at wavelengths of 50 \um\ and 250 \um\ as well as interferometric observations by ALMA  for the dust-continuum at 158 \um. The images are produced by taking into account the noise as well as the PSF (for mock imaging) or beam sizes (for ALMA configuration in CASA). We analysed our galaxies for various SNR $\in [5,20]$. To evaluate the sizes of galaxies in the simulation, after radiative transfer, we used a curve of growth method, by using circular apertures of increasing radius from the most bound particle and interpolating the radius which encloses 50 percent of the light. We used \texttt{statmorph} \citep{10.1093/mnras/sty3345} to evaluate the sizes from the mock observations. Using both these methods we are able to compare the sizes of mock observations with intrinsic sizes. 
Using these methods, we also calculate the sizes in IR regime. 

We compare the slope parameter in the power laws (Equations \ref{eqn:lumpowerlaw} \& \ref{eqn:Power Law}) of size evolution with cosmic time as well as size evolution against luminosity at fixed redshift, for both simulations and observations. We also analyse how the sizes vary with increasing mass and wavelength of observations in both UV and IR. Our main findings from these studies are as follows:
   \begin{enumerate}
      \item Mock NIRCam observations of \flares\ galaxies at rest frame FUV (1500 \AA) show a decreasing size evolution with redshift, which is consistent with many observational studies.  For observed sizes, all our sample lies in the range  m=$1.0-2.0$ as predicted in \citet{2018MNRAS.477..219M}. Our sample are in agreement with observational studies at both low luminosities \citep{Oesch_2010,Shibuya_2015,Ono_2013,Kawamata_2018} and high luminosities \citep{Holwerda_2015}. Our intrinsic size match the trends of dust attenuated size evolution in different luminosity bins as presented in \citetalias{10.1093/mnras/stac1368}.
      \item For fixed redshifts, the size-luminosity evolution of the mock observations agree with \citet{Shibuya_2015,Holwerda_2015,Grazian,Bouwens_2022} study showing a positive $\beta \approx 0.27$  (at $z=7$). The slopes are also consistent with previous dust-attenuated simulation studies (\citetalias{10.1093/mnras/stac1368}, \citet{10.1093/mnras/stac380}).  Mock observed slopes are slightly shallower compared to early \jwst\  \citep{Yang22} rest-frame UV size-luminosity slopes from F150W filter although at z>6 we are able to match the size-luminosity scatter from the recent \jwst\ surveys well. We also find a broadly non-evolving $\beta$ with increasing redshift similar to \citet{Shibuya_2015}. The intrinsic size $\beta$ which can recover the dispersed sample reports a $\beta=0.59$.

      \item Due to observational effects of noise and PSF, at lower SNR, the mock observational sizes are underestimated. While at higher SNR due to the reduced effect of noise sizes are underestimated at lower redshift (z = 5, 6, 7, 8) in NIRCam observations and at higher redshifts (z = 9, 10) the sizes tend to be slightly overestimated where the effect of PSF leads to bright cores becoming bigger. Hence, the ratio of mock observation sizes to intrinsic sizes is a function of signal to noise ratio.
      
      \item Mock photometric IR observations also show the effect of noise and PSF as the photometric sizes are underestimated compared to intrinsic sizes. For ALMA simulations at High resolution ALMA ($\approx 0.01''$), the low-surface brightness extended emissions of a galaxy will be lost leading to underestimation of sizes with a factor=0.89 times intrinsic size. Our sizes and achieved SNRs for the ALMA low angular resolution ($\approx 0.3''$) match with the recent studies \citet{mitsuhashi2023almacristal,ikeda2024almacristalsurveyspatialextent}. Due to beam effects the observed low angular resolution sizes are nearly 10 times the intrinsic size of the galaxy. Nearly 50 percent of the galaxy at this resolution fall below the beam size. Low SNR noise near the signal in these low-resolution ALMA observations ($\approx 0.3''$)  can also act as extended emissions, hence also leading to further extended sizes. The sizes in IR spectrum follow the same power law as UV but the sizes are substantially smaller for very high-resolution imaging. As the dust sizes remain compact at these redshift power laws are less steep with slope m = 0.24 for photometric observations, m = 0.16 for ALMA observations and m = 1.39 for intrinsic sizes.
      
      \item The ratio of intrinsic sizes in UV and IR increases with increasing mass, which is due to higher mass galaxies having higher star formation rates and being dominated by young blue stars. Low mass galaxies (log$_{10}(\text{M}^*/\textup{M}_\odot) \approx 8.8-9.8$) at $z<6$ show a sudden dip in 
      UV-IR size ratio due as this sample also contains galaxies transitioning from dispersed to compact star formation as described in \citetalias{roper2023flares}. The observed UV-IR size ratio (for simulated NIRCAM and ALMA imaging), shows dust sizes to be greater than UV sizes agreeing with recent studies \citep{2024A&A...686A.187P,2022ApJ...934..144F,mitsuhashi2023almacristal}.
   \end{enumerate}

The variation of observed from intrinsic sizes highlights the importance of accounting for observational effects such as  PSF and SNR when interpreting galaxy size measurements. As higher resolution photometry of galaxies in the EoR become increasingly available with \jwst, the constraints on the size evolution slopes will become increasingly robust in UV.  Observational effects such as SNR and angular resolution also affect the observed ALMA sizes. Studying sizes in IR to a high accuracy will also require very high-resolution observations which can resolve both small scale structures and low surface brightness extended emissions. Combining multi-wavelength studies of galaxy sizes in IR and UV can help us unravel the physics beneath dust-obscured star formation to understand the interplay of star formation, chemical enrichment and feedback.

A proper comparison between observations and theory becomes paramount to, validate, and refine our understanding of galaxy evolution. Next generation simulations must aim to not only reproduce the UV luminosity functions, but also predict the observed size and luminosity distribution of galaxies across different observed wavelengths. In the upcoming future, new instruments like Extremely Large Telescope (ELT) and Very Large Array (VLA) will be coming online. Investigations of the morphology and size evolution of high redshift galaxies will be made possible by the unmatched spatial resolution and sensitivity that ELT and VLA will offer in the optical, infrared, and radio wavelengths. Square Kilometre Array (SKA) also coming online, will enable studies of the cold gas reservoirs and large-scale structures around these galaxies, hence analysing  environmental factors behind galaxy growth. Together, these instruments will help develop a comprehensive, multi-wavelength view of early galaxy evolution.

Coupling these next-generation multi-wavelength observations with simulations will refine our knowledge of galaxy formation and evolution at the epoch of re-ionisation, providing us with deeper insights into the complex interplay of gas, dust and radiation. 
\section*{Data availability statement}
The size analysis data and the code to reproduce the plots is available at \href{https://github.com/paurush-p/FLARES_XVI_plots}{https://github.com/paurush-p/FLARES\_XVI\_plots}.

\begin{acknowledgements}
      This work has been performed using the Danish National Life Science Supercomputing Center, Computerome. This work also used the DiRAC@Durham facility managed by the Institute for Computational Cosmology on behalf of the STFC DiRAC HPC Facility (www.dirac.ac.uk). The equipment was funded by BEIS capital funding via STFC capital grants ST/K00042X/1, ST/P002293/1, ST/R002371/1 and ST/S002502/1, Durham University and STFC operations grant ST/R000832/1. DiRAC is part of the National e-Infrastructure. The Cosmic Dawn Center (DAWN) is funded by the Danish National Research Foundation under grant DNRF140.
      
      We also wish to acknowledge the following open source software packages used in the analysis: NUMPY \citep{Harris_2020}, SCIPY \citep{Virtanen_2020}, PANDAS \citep{reback2020pandas}, ASTROPY \citep{astropy:2013,astropy:2018,astropy:2022}, PILLOW \citep{clark2015pillow}, MATPLOTLIB  \citep{4160265} and STATMORPH \citep{10.1093/mnras/sty3345}. PP would also like to thank Dr. Anurag Nishad and Dr. Anumpama Sharma  for guidance towards completion of his thesis leading to this study. The \flares\ team would like to acknowledge the enormous contributions of the late Prof. Peter Thomas to the \flares\ project; we will miss his guidance and great wit. We also thank the anonymous referee for their helpful comments.
\end{acknowledgements}


\bibpunct{(}{)}{;}{a}{}{,} % to follow the A&A style

% for the bibliography, at the end
\bibliographystyle{aa} % style aa.bst
\bibliography{biblio} % your references Yourfile.bib

\begin{appendix}
%\section{Comparison of size determination with previous FLARES studies}\label{sec.sizecomp}

\section{Size calculation convergence test}\label{converge}
\begin{figure}[h!]
	\centering
		\includegraphics[width=7.5cm,keepaspectratio]{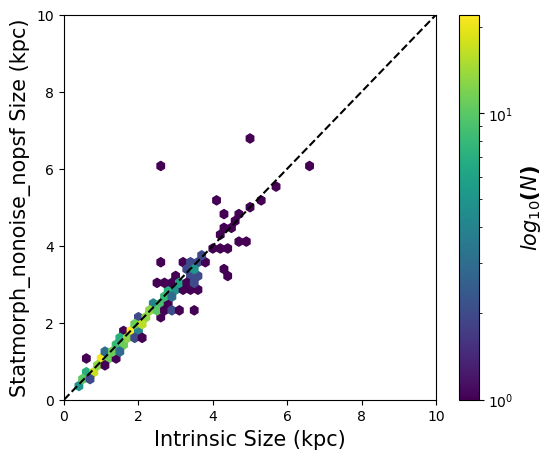}
	\caption[]{We plot the sizes calculated by \texttt{statmorph} for noise and PSF free radiative transfer output on y-axis and the respective intrinsic sizes calculated by the Iterative Aperture Method on the x-axis. The black dotted line represents the 1:1 relation with the colourbar showing the number density in the hexbins. }
	\label{fig:sizecon}
\end{figure}

As described in the text we have used two separate methods to calculate intrinsic galaxy sizes and observed sizes and hence we tested the convergence of sizes using both methods on noise and PSF free radiative transfer output. We created a subsample consisting of 626 galaxies which follow the same relative distribution of galaxy stellar mass and redshift as in the larger sample. We then calculated sizes using \texttt{statmorph} while removing cases of segmentation failure from the sample. The resulting size relation is presented in Figure \ref{fig:sizecon}

\section{NIRCam calculated resolutions}

The resolution used for NIRCam based on filters and redshift, calculated by \citep{astropy:2013,astropy:2018,astropy:2022} is presented in Table \ref{tab:resolution}.

\begin{table}[H]
    \centering
    \begin{tabular}{cc}
    \hline
        \textbf{Redshift(z)} & \textbf{Resolution(pxs by pxs)}\\ \hline \hline
        5 & 302x302 \\ 
        6 & 332x332 \\ 
        7 & 363x363 \\ 
        8 & 393x393 \\
        9 & 425x425 \\
        10 & 456x456 \\\hline
    \end{tabular}
    \caption{NIRCam Image Resolution at different redshifts for a field of view of 60 pkpcx60 pkpc (size of radiative transfer simulations)}\label{tab:resolution}
\end{table}

The radiative transfer outputs are regridded to these resolutions for mocking observations.
\section{Effect of dust on luminosity}\label{sec.dust}
\begin{figure}[]
	\centering
		\includegraphics[width=\linewidth,height=6cm,keepaspectratio]{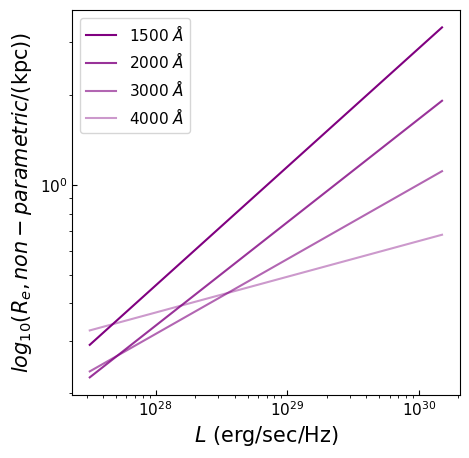}
	\caption[]{Intrinsic non-parametric sizes (y-axis) are fit for the size luminosity relation at z=7 for 1500 \AA, 2000 \AA, 3000 \AA\ and 4000 \AA\ respectively, which is plotted against the luminosities at the respective wavelengths (x-axis). The slopes decrease with increasing wavelength.}
	\label{fig:dusteffect}
\end{figure}

To evaluate the evolution of attenuation effect of dust with wavelength we fit the size-luminosity relation at z=7 at 1500 \AA, 2000 \AA, 3000 \AA\ and 4000 \AA\ for the intrinsic non parametric sizes as defined in \S \ref{nonparam} in Figure \ref{fig:dusteffect}. We used non-parametric sizes to prevent any clumpy or dispersed morphological shapes effecting the size-luminosity fits, allowing us to probe the effect of dust independently. As the attenuation effect of dust weakens with increasing wavelength, we see that at a fixed redshift the slopes of the luminosity size fit decreases.
\section{Size-offset relation to galaxy characteristics}
\begin{figure}[h!]
	\centering
		\includegraphics[width=7.5cm,keepaspectratio]{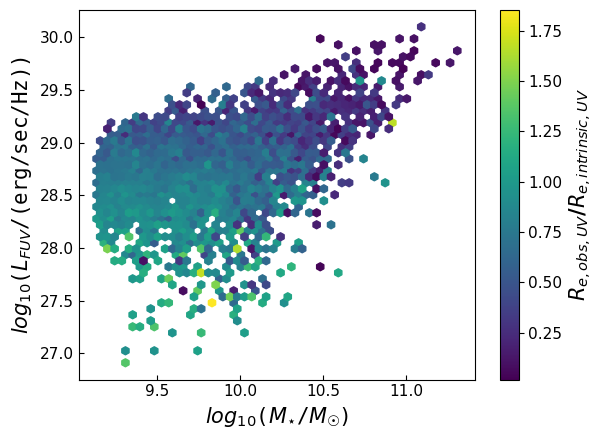}
	\caption{Ratio of simulation to observational sizes is plotted in the plane of stellar mass (x-axis) and FUV luminosity (y-axis) for galaxies at z=5}
	\label{fig:sizedep1}
\end{figure}

To determine why there is an offset between observational sizes and intrinsic sizes, in Figure \ref{fig:sizedep1}, we plot the ratio of sizes in simulations to observational size on a plane of FUV luminosity and stellar mass, seeing no significant correlation. 

\section{Effect of noise and PSF on r20 and r80}\label{sec:noisefield}
\begin{figure}[]
\centering
   \includegraphics[width=7.5cm]{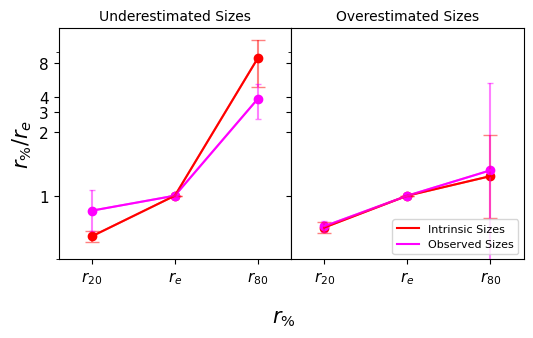}
     \caption{Ratio of r20 and r80 with the half-light radius (y-axis) are plotted for galaxies with underestimated sizes at SNR=5 and overestimated sizes at SNR=20. }
     \label{fig:r20r80}
\end{figure}

We analysed how the relative distribution of light for the galaxies changes due to observational effect. Figure \ref{fig:r20r80} evaluates r20 \& r80 of galaxies with size underestimation at SNR=5 and overestimation at SNR=20 for intrinsic and observed imaging. All sizes are normalised by the half-light radius for that specific imaging.
When underestimation takes place at low SNR, the observed r20 shows a subsequent increase signifying that low luminous regions blend with noise leading to a loss in luminosity which is not recovered by PSF enlarging the cores. This is compensated by the decrease in observed r80. At high SNR, the loss in luminosity by noise is less and which is also recovered by PSF's enlargement of the bright cores. This relative shift of luminosity to the centre leads to only a slight increase in observed r80, with observed r20 being close to intrinsic r20.

\section{Simulated observations in a singular noise field}\label{sec:noisefield}
\begin{figure}[]
\centering
   \includegraphics[width=7cm]{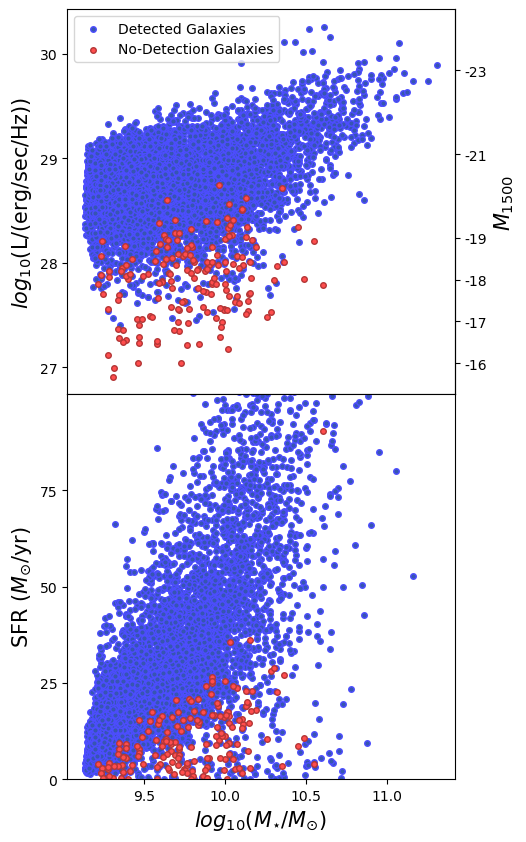}
     \caption{Upper panel: FUV luminosities (y-axis) are plotted against their respective Stellar Masses (x-axis). Detected galaxies are shown in blue and non-detections in red, at SNR=5. The scatter is of the whole sample across all redshifts with most non-detections lying at z=5,6. Lower Panel: Similar figure as the upper panel, we now compare SFR (y-axis) of the galaxies detected using mock images and the ones which were not at different Stellar Masses (x-axis) at SNR=5. Detected galaxies are shown in blue and non-detections in red.}
     \label{fig:nodetectlum}
\end{figure}

We analysed the galaxies when they are placed in a singular noise field to analyse the effect of same observation time on detection and non-detection. We found that out of a total sample of 6890 galaxies 167 will not be detected at  1500 \AA\ 
by \textit{JWST}.
The galaxies which are not detected have low luminosities compared to the galaxies which were detected for similar stellar mass (See Figure \ref{fig:nodetectlum}). On further visual inspection of the galaxy images, we see that these galaxies are very dispersed and hence at lower SNR, a lot of the detail in these galaxies falls below the noise depth. These galaxies also have a comparatively lower SFR (See Figure \ref{fig:nodetectlum}) to the rest of the data-set. The IR sizes of these galaxies are compact and similar to other galaxies which are detected. The infrared characteristics as well as example images of these galaxies is discussed in Appendix.

The galaxies which are not detected in UV spectrum, as discussed earlier, have a comparatively lower SFR compared to the general population and they have a low luminosity detected at 1500 \AA, but these galaxies have similar luminosities (to general population) in Infrared signifying older stellar population which is also responsible for the stellar mass being comparable to the general population.

\begin{figure}[h!]
\centering
   \includegraphics[width=7cm]{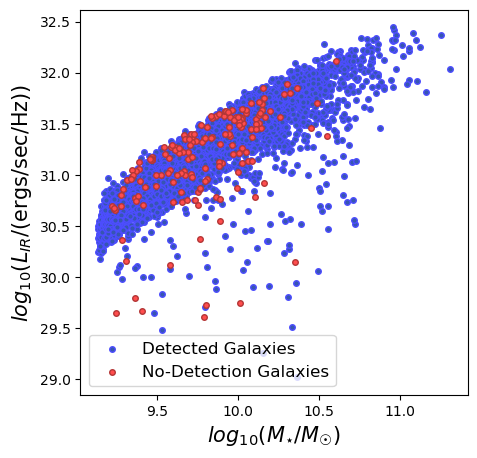}
     \caption{Infrared Luminosity at 250 $\mu$m (y-axis) is plotted against Stellar Mass (x-axis) for the whole population to distinguish infrared properties between FUV detected galaxies against the non-detected ones. The non-detections show a similar spread and pattern to detections.}
     \label{fig:irnodetect}
     
\end{figure}

We also look at the IRX-$\beta$ plot to see if this non-detection can also be due to presence of dust but the galaxies which are not detected are spread across a wide value of IRX and $\beta$.

\begin{figure}[h!]
\centering
   \includegraphics[width=7cm]{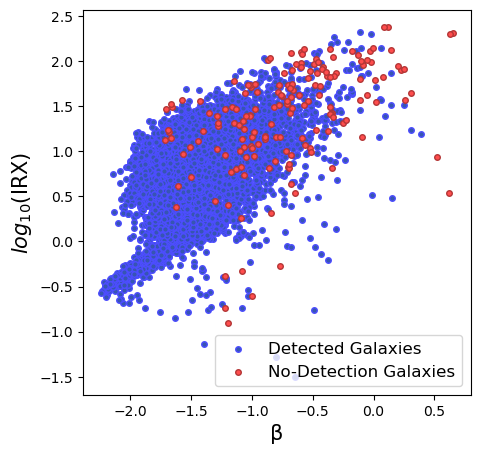}
     \caption{IRX (y-axis) - $\beta$ (x-axis) is plotted for the whole sample to investigate the effect of dust between FUV detected galaxies against the non-detected ones. No clear pattern for non-detections is established to allude to dust causing non-detections. }
     \label{fig:irxnodetect}
     
\end{figure}
\begin{figure}[h!]
\centering
   \includegraphics[width=8cm]{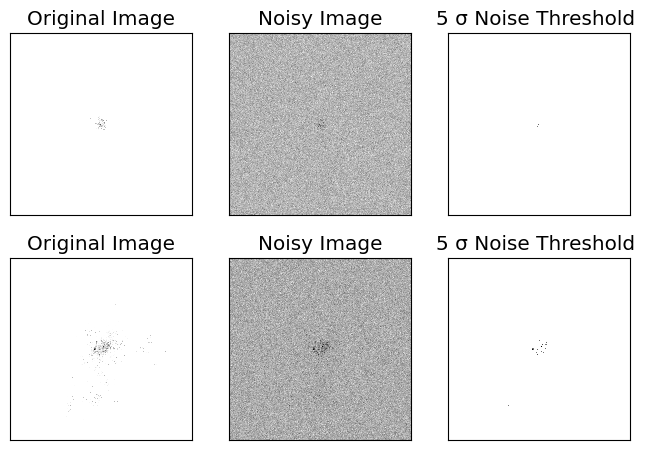}
     \caption{Sample no-detection galaxies at various stages of image analysis is shown above. The first panel shows the output of radiative transfer, the second panel shows the simulated noisy image with respect to NIRCam and the last panel shows the residual image obtained after a 5-sigma threshold is applied to remove noise. The final images lead to segmentation algorithm finding no credible sources.}
     \label{fig:irxnodetect1}
     
\end{figure}

A few sample images of these non-detected galaxies with the original, noisy image and then the resultant image when a 5 $\sigma$ noise threshold is applied are given in Figure \ref{fig:irxnodetect1}.

\section{Effect of noise on ALMA imaging}

\begin{figure}[h!]
\centering
   \includegraphics[width=7.5cm]{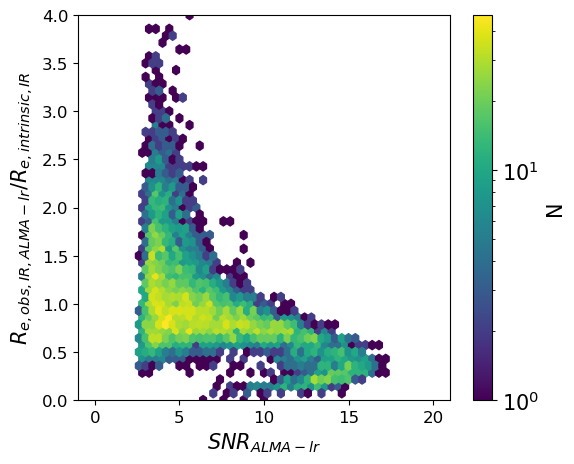}
     \caption{The figure shows observed to intrinsic size ratio at 158 \um\ (y-axis) plotted against the observed signal to noise ratios(x-axis). The spread of galaxies towards higher size ratios increases with decreasing SNR.}
     \label{fig:snreffect}
     
\end{figure}

At low signal to noise ratios in ALMA with very low resolution, noise around the bright centres of galaxies can appear as extended emission. To analyse this effect, in Figure \ref{fig:snreffect} we plotted the 158\um\ observed to intrinsic size ratio against observed SNR levels. We see that with decreasing SNR the spread of the size ratio increases towards higher values. This effect is opposite of what is seen for UV emission where the decreasing effect of noise leads to mock \jwst\ sizes getting closer to intrinsic sizes.

\onecolumn
\section{Mock high angular resolution ALMA imaging}
\begin{figure*}[h!]
\centering
   \includegraphics[width=17.5cm]{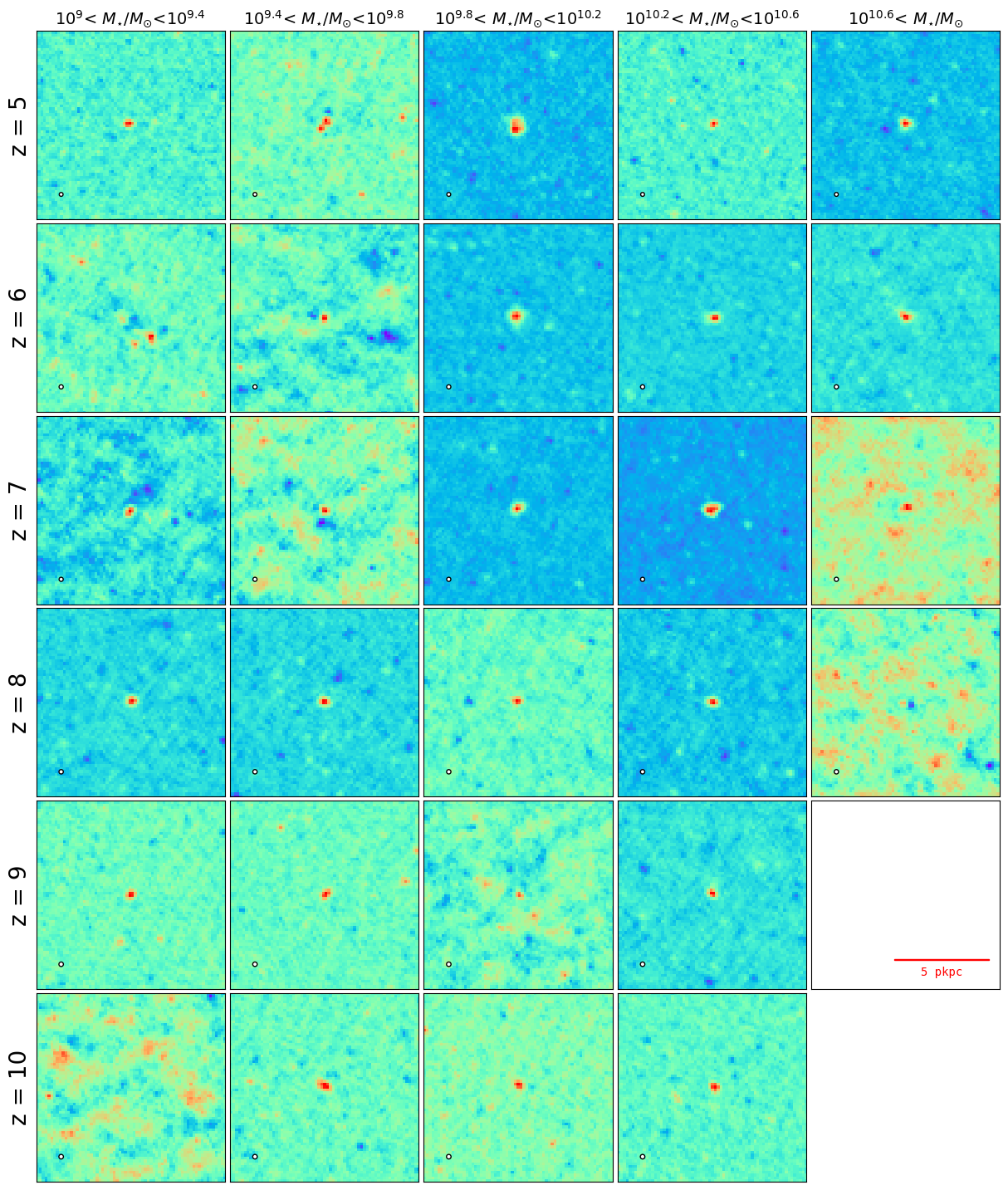}
     \caption{The figure shows sample galaxies (same galaxies as in Figure \ref{fig:NIRCamfalse}) at different redshifts and various mass bins as observed by ALMA at 158 \um\ at very high resolution ($\approx 0.02"$) using C8 configuration. The aperture of the images is 10 pkpc. The beam size is shown by the circle at bottom left corner}
     \label{fig:ALMAfalsehighres}
     
\end{figure*}

\begin{figure*}[h!]
\centering
   \includegraphics[width=17.5cm]{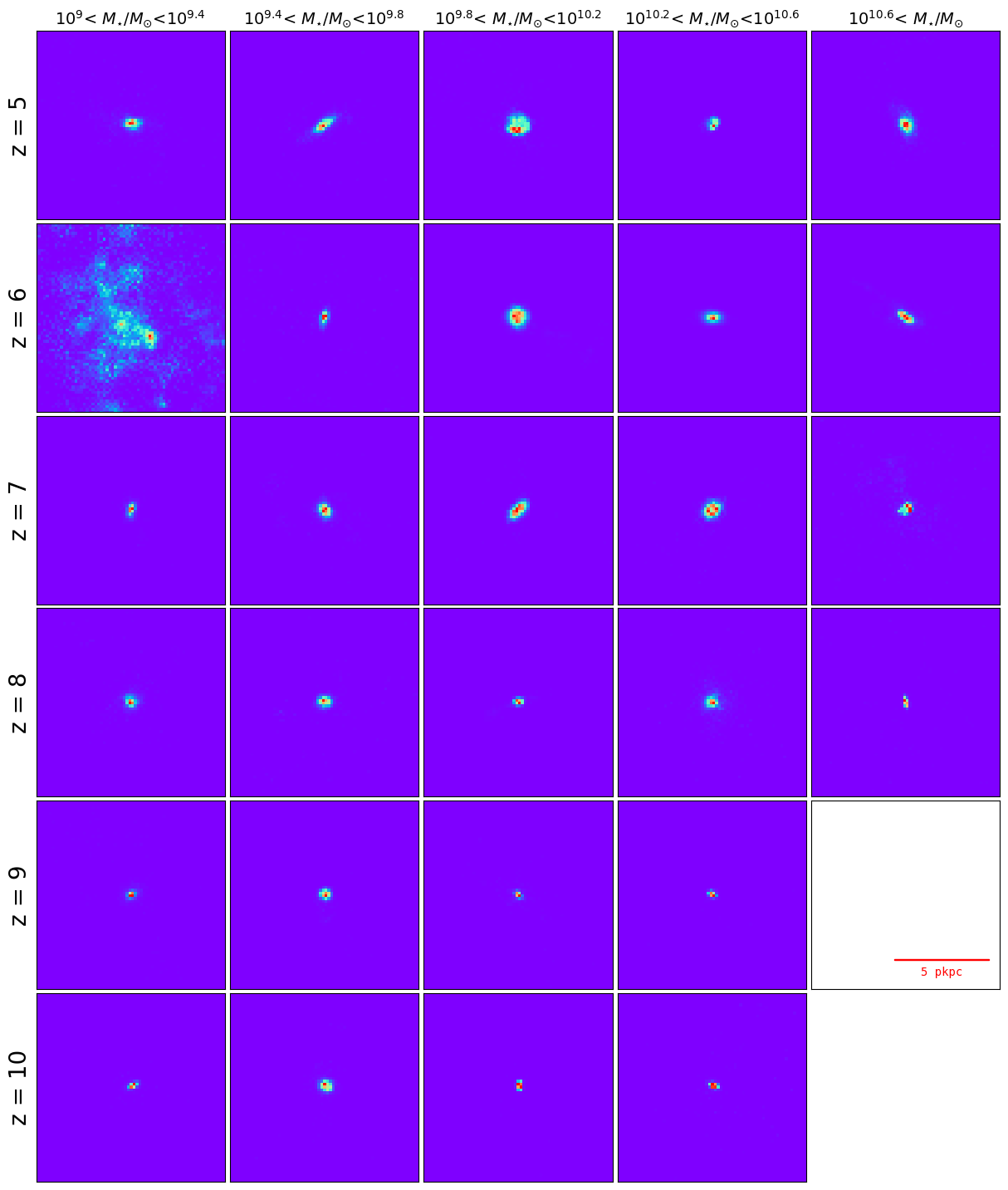}
     \caption{The figure shows the radiative transfer output at 158 \um\ of the sample galaxies (same galaxies as in Figure \ref{fig:NIRCamfalse}). The aperture of the images is 10 pkpc.}
     \label{fig:realIR}
     
\end{figure*}
Theoretically ALMA can go to very high resolutions, even beyond JWST resolutions, at a lot of galaxy structures will be resolved. The observational time required to achieve these resolutions on a single source remains unjustifiable, but to look at the effects of resolution on ALMA sizes we simulated galaxies' observations with C8 configuration as described in main text. The imaging is presented in Figure \ref{fig:ALMAfalsehighres}.

\section{Size-luminosity relation comparison}

\begin{figure*}[h!]
\centering
   \includegraphics[width=17.5cm]{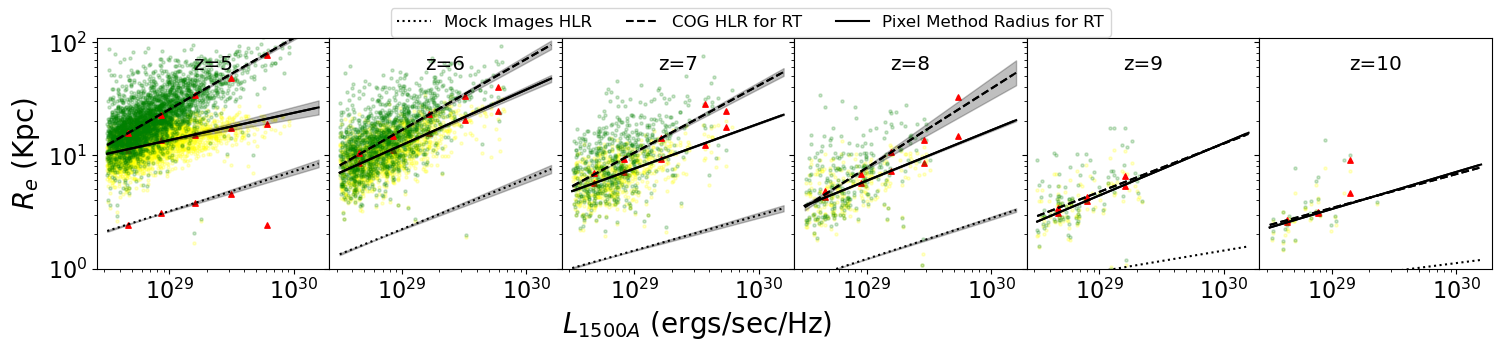}
     \caption{Non-weighted luminosity size relation evaluated  at different redshifts is shown above. The y-axis shows the UV 1500 \AA\ sizes in kpc and the x-axis shows luminosity at the same wavelength. The green scatter shows our sample of galaxies analysed with radiative transfer for Curves of Growth sizes using circular apertures whereas the yellow scatter shows the sizes for non-parametric Pixel Method as presented in \citetalias{10.1093/mnras/stac1368}.}
     \label{fig:lscm}
\end{figure*}

To show the effect of using a non-parametric pixel based method as presented in \citetalias{10.1093/mnras/stac1368} to this studies' curves of growth method we fit the size luminosity relation for both size calculation methods to the radiative transfer output in Figure \ref{fig:lscm}.

\section{Size overestimation effect of PSF}\label{sec:noisefield}
\begin{figure*}[h!]
\centering
   \includegraphics[width=17.5cm]{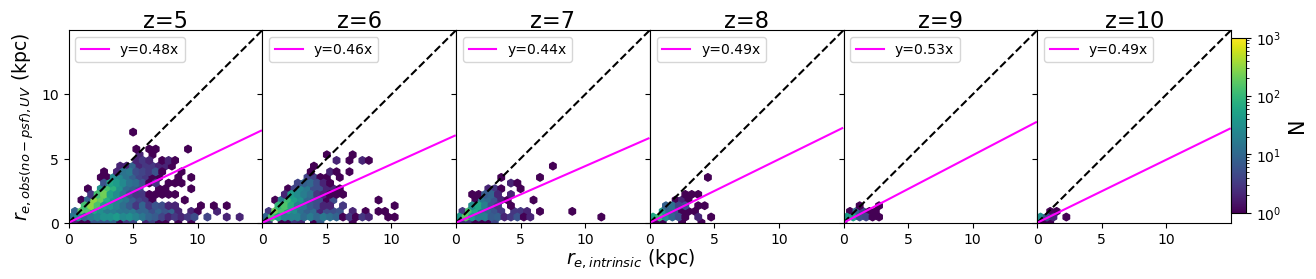}
     \caption{Observed UV sizes with no PSF effect from mock images (y-axis) are plotted as a function of intrinsic sizes (x-axis) at SNR=15. The black dashed line is a 1:1
relation whereas the magenta line is the best fit y=mx relation where y is mock image effective radius and x is simulation effective radius}
     \label{fig:psfeffect}
\end{figure*}

To understand the effect of PSF independent of noise we also simulated UV 1500 \AA\ imaging without effect of PSF but Gaussian noise included. We did this at SNR=15 as mock images with full observational effects show both underestimation and overestimation for this sample. We also removed any galaxy with greater than 5 segments detected, to avoid extremely clumpy systems as these clumps without the effect of PSF will be extremely small and isolated. The comparison of intrinsic sizes to the no-PSF imaging sizes is present in Figure \ref{fig:psfeffect}. We see that at all redshifts most of the observed sample is underestimated in size with factors of 0.48-0.53 times the intrinsic size. For imaging with PSF as shown in Table \ref{tab:compslope} the underestimation is seen to reduce from 0.64 times at z=5 to 0.83 times at z=8 and then overestimation takes place with sizes being 1.04-1.07 times intrinsic sizes at z=9, 10.
\begin{figure}[]
\centering
   \includegraphics[width=7.5cm]{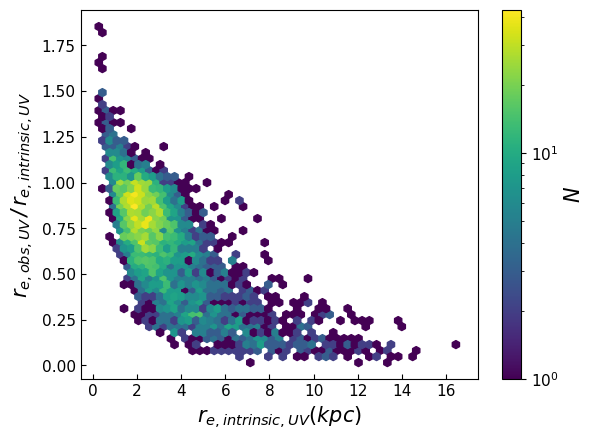}
     \caption{Observed to Intrinsic UV size ratio (y-axis) is plotted against the intrinsic sizes (x-axis) at SNR=5 and z=5. }
     \label{fig:compact}
\end{figure}
It is clear that PSF causes overestimation or restricts underestimation of sizes due to noise. As the PSF spreads light from bright sources, sizes of more compact systems should be affected by this spread more than dispersed systems. In Figure \ref{fig:compact} we plotted the ratio of intrinsic and observed sizes against the intrinsic sizes and found that the more compact a galaxy is the less underestimated its observed sizes will be and can even be overestimated. 

\section{Very high SNR UV imaging}\label{sec:highsnr}
\begin{figure*}[h!]
\centering
   \includegraphics[width=17.5cm]{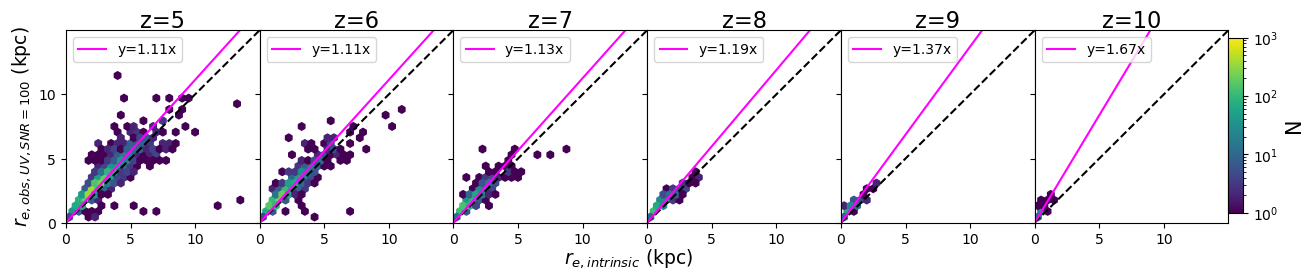}
     \caption{Observed UV sizes at SNR=100 mock images (y-axis) are plotted as a function of intrinsic sizes (x-axis). The black dashed line is a 1:1
relation whereas the magenta line is the best fit y=mx relation where y is mock image effective radius and x is simulation effective radius}
     \label{fig:snr100}
\end{figure*}

\begin{figure*}[h!]
\centering
   \includegraphics[width=14cm]{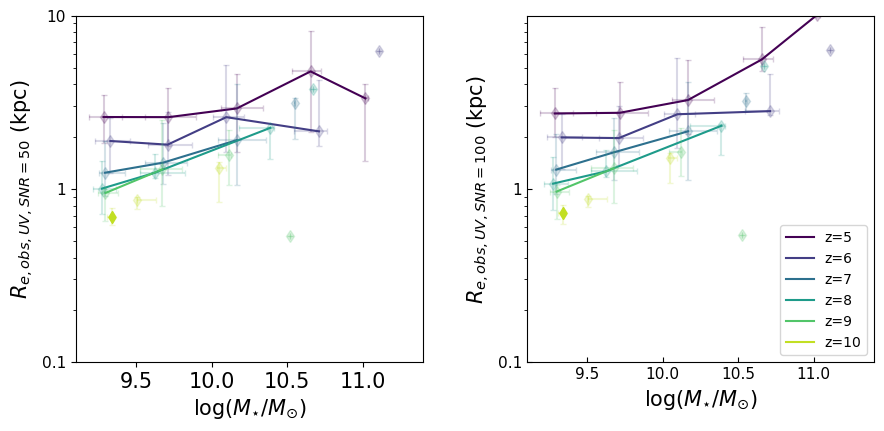}
     \caption{Left panel: The evolution of the mock observation sizes at SNR=50 (y-
axis, in kpc) as function of stellar mass (x-axis) for $ z \in [5, 10]$. Right panel: Evolution of sizes from mock observations at SNR=100 as function of stellar
mass.}
     \label{fig:masssnr100}
\end{figure*}

To further verify our overestimation being the effect of observational effects, in Figure \ref{fig:snr100} we plotted the relation between intrinsic and galaxies at SNR=50 and SNR=100 with the PSF applied. As the effect of noise becomes minimal at eating away at the faint parts of the galaxy, we see an overestimation of size at all redshifts. In Figure \ref{fig:masssnr100} we also plotted the Stellar mass-size relation at these SNRs, seeing the evolution of sizes following the same positive trend as intrinsic-size relation. 

\end{appendix}

\end{document}